\documentclass[journal]{IEEEtran} %

\usepackage[dvipdfmx]{graphicx}          %
\usepackage{here,amsmath,amssymb,bm,arydshln,cite}

\ifCLASSINFOpdf
\else
\fi
\usepackage{url}
\usepackage{fancyhdr}

\fancypagestyle{firstpage}{\lfoot{\scriptsize \copyright 2015 IEEE. Personal use of this material is permitted. Permission from IEEE must be obtained for all other uses,
in any current or future media, including reprinting/republishing this material for advertising or promotional purposes, creating new collective works, for resale
or redistribution to servers or lists, or reuse of any copyrighted component of this work in other works. The final version of record is available at 
\url{http://dx.doi.org/10.1109/TMBMC.2015.2500567}} \cfoot{}}

\begin{document}

\title{Coordinated Spatial Pattern Formation in Biomolecular Communication Networks}

\author{Yutaka~Hori,~\IEEEmembership{Member,~IEEE,}
        Hiroki~Miyazako,  %
				Soichiro Kumagai,
        and~Shinji~Hara,~\IEEEmembership{Fellow,~IEEE}%
\thanks{Y. Hori is with Department of Computing and Mathematical
        Sciences, California Institute of Technology, CA 91125
        USA. {\tt yhori@caltech.edu}.}%
\thanks{H. Miyazako, S. Kuamgai and S. Hara are with Department of Information
        Science and Technology, The University of Tokyo, Tokyo 
        113-8656 Japan. {\tt \{Hiroki\_Miyazako,
        Shinji\_Hara\}@ipc.i.u-tokyo.ac.jp}. {\tt kumagai@hapis.k.u-tokyo.ac.jp}}.%
\thanks{Y. Hori was supported by JSPS Fellowship for Research
        Abroad. This work was supported, in part, by Grant-in-Aid for Scientific Research (A) of the Ministry of Education, 
Culture, Sports, Science, and Technology, Japan, No. 21246067, and Grant-in-Aid for
JSPS Fellows under grant number 15J09841.}}

\maketitle

\thispagestyle{firstpage}
\begin{abstract}
This paper proposes a control theoretic framework to model and
 analyze the self-organized pattern formation of molecular concentrations 
in biomolecular communication networks, emerging applications in
 synthetic biology. In biomolecular communication networks, bio-nanomachines, or
 biological cells, communicate with each other using a cell-to-cell
 communication mechanism mediated by a diffusible signaling molecule,
 thereby the dynamics of molecular concentrations are approximately
 modeled as a reaction-diffusion system with a single diffuser.
We first introduce a feedback model representation of the
 reaction-diffusion system and provide a systematic local
 stability/instability analysis tool 
using the root locus of the feedback system.
The instability analysis then allows us to analytically derive the conditions for the self-organized
 spatial pattern formation, or Turing pattern formation, of the
 bio-nanomachines. We propose a novel synthetic biocircuit motif called
 activator-repressor-diffuser system and show that it is one of 
 the minimum biomolecular circuits that admit self-organized patterns
 over cell population.
\end{abstract}


\IEEEpeerreviewmaketitle

\section{Introduction}
Designing the coordinated dynamical behavior of cell populations is 
one of the important challenges in synthetic biology. 
The study of cell population control is expected to provide not only 
the better understanding of biology but also the ability to build 
biomimetic nanomachines, or bio-nanomachines, for bioengineering
applications such as tissue engineering and targeted drug delivery
 \cite{Akyildiz2012, Nakano2014}. 
In the last decade, researchers designed 
a variety of biomolecular circuits that perform population-level behavior such as 
synchronized oscillations \cite{Danino2010}, population control \cite{You2004,Balagadde2008},
and multicellular computation \cite{Tabor2009,Tamsir2011,Regot2011}. 
In these works, intercellular coordination was enabled by a
cell-to-cell communication mechanism called quorum sensing, where cells
communicate with each other by releasing and receiving diffusible signaling molecules.

\par
\smallskip
One of the earliest synthetic biocircuits for spatial pattern formation
of cell population was demonstrated in Basu {\it et al.} \cite{Basu2005},
where sender cells transmit a quorum sensing autoinducer molecule and
receiver cells respond by expressing fluorescent protein to form 
predefined spatial patterns.
This work was later followed by Liu {\it et al.} \cite{Liu2011}, where another
biomolecular circuit was developed to form a periodic pattern of cell density in a self-organized manner. 
From an engineering point of view, self-organized pattern formation is
preferable since it does not require extra design steps of external
inputs and spatial allocation of cells. 
Moreover, the study of self-organization is important in developmental
biology to understand morphogenesis \cite{Koch1994,Kondo2010}.

\par
\smallskip
Recently, Hsia {\it et al.} \cite{Hsia2012} introduced a dynamical
model of biomolecular communication networks using reaction-diffusion
equations, where diffusion-based cell-to-cell communication was modeled
based on the Fick's laws of diffusion \cite{Edelstein-Keshet2005}. 
It was then shown that a biomolecular communication network can form 
a self-organized concentration gradient of chemical species over cell populations
despite the averaging effect of molecular diffusion. 
Such self-organized spatio-temporal pattern is known as Turing pattern
\cite{Turing1952}. 

\par
\smallskip
Although the theory of Turing pattern formation was extensively studied
over the last 60 years \cite{Gierer1972,Edelstein-Keshet2005},
many existing works of Turing pattern formation, which are often applied
to developmental biology (see \cite{Othmer2009} for example), are not directly
applicable to today's engineering applications in biomolecular communication networks because of the
difference of the number of diffusible molecules. 
Specifically, in biomolecular communication networks, 
only a single molecular species, or a small signaling molecule, can
diffuse between cells and most proteins and mRNAs are localized in a cell, while
the classical works assume that all molecular species can diffuse in the
medium. 
Hence, it is desirable to develop a novel theoretical framework that
specifically targets the systematic modeling and analysis of the spatio-temporal dynamics of 
biomolecular communication networks with a single diffusible molecule.

\par
\smallskip
In this paper, we consider a class of reaction-diffusion systems that
approximately models biomolecular communication networks, emerging
applications in synthetic biology.
Our interest here is to study the structural properties of 
the systems that come from the physical constraints of biomolecular
communication and utilize them to derive an analysis framework 
rather than to study general reaction-diffusion systems.
Specifically, the goals of this paper are (i) to present a stability
analysis tool for biomolecular communication network systems
 and (ii) to use it to study the structures of biocircuits that are
 required for Turing pattern formation over the population of cells. 
From a synthetic biology viewpoint, it is useful to characterize the
required structures of biocircuits as it specifies the design space of possible
combinations of genetic parts. 

\par
\smallskip
We first present that the local stability analysis of the 
reaction-diffusion systems boils down to a simple graphical test using
the root locus method \cite{Levine1996}. 
We show that the root locus is characterized by the dynamics of
biocircuits inside cells and the complete circuit including diffusers,
which allows us to provide physical interpretations of our theoretical results. 
Using the stability/instability analysis tool, we study conditions for Turing pattern formation 
in synthetic biomolecular communication networks.
In particular, we mathematically prove that the minimum biocircuit
producing coordinated spatial patterns is composed of three genes, and moreover, 
certain activation-repression network patterns are necessary. 
As an example of the three-component biomolecular circuits, we propose
activator-repressor-diffusor motif and show that a 
synthetic biocircuit with this motif can admit Turing instability using
illustrative numerical simulations.

\par
\smallskip
The organization of this paper is as follows. 
In Section~\ref{model-sec}, we provide a control theoretic modeling
framework for biomolecular communication networks and introduce a useful 
decomposition that simplifies the stability analysis of the system.
Then, in Section~\ref{conttheo-sec},  we provide a systematic
stability analysis tool based on the root locus method and demonstrate on a novel 
activator-repressor-diffusor biomolecular circuit.
Section~\ref{turing-sec} is devoted to deriving the condition for Turing
pattern formation. %
Then, in Section \ref{threecomp-sec}, the class of reaction networks that can admit
Turing patterns is characterized. Finally, Section \ref{conc-sec} concludes the paper. 

\par
\smallskip
The authors' conference papers \cite{MiyazakoCDC,HoriSICE2014} illustrated 
the outlines of the Turing instability analysis method shown in this
paper, and the proofs were partly presented in the paper written in
Japanese \cite{Miyazako2013}. 
In contrast with these papers, the present paper emphasizes emerging applications in synthetic biology. 
For this purpose, all results are presented in a two dimensional spatial
domain. In particular, a novel activator-repressor-diffuser biomolecular circuit motif is proposed as an example of minimum
biocircuits to achieve spatially coordinated pattern formation. 
Biological relevance and challenges for implementation are also
discussed. 
In addition, we provide a formal mathematical statement of our key
result, a necessary and sufficient exponential stability condition (Proposition 1), for the first
time with a complete proof, then the relation with the existing theoretical result \cite{Arcak2011}
is discussed. Other mathematical results are also shown with an updated
rigorous assumption (Assumption 1) and complete proofs.

\par
\smallskip
The following notations are used throughout this paper: 
$\mathbb{N}_{0} := \{0,1,2,\cdots\}$. 
$\mathbb{R}_{0+} := \{r \in \mathbb{R}~|~r \ge 0\}$.
$\mathbb{C}_{+} := \{c \in \mathbb{C}~|~\mathrm{Re}[c] > 0\}$.
$\mathbb{C}_{0+} := \{c \in \mathbb{C}~|~\mathrm{Re}[c] \ge 0\}$.
$\mathbb{C}_{-} := \{c \in \mathbb{C}~|~\mathrm{Re}[c] < 0\}$.
$I_n$ is an $n$ by $n$ identity matrix.

\section{Modeling of Biomolecular Communication Networks}
\label{model-sec}

\subsection{Model description}
\begin{figure}[tb]
\centering
\includegraphics[clip,width=6.0cm]{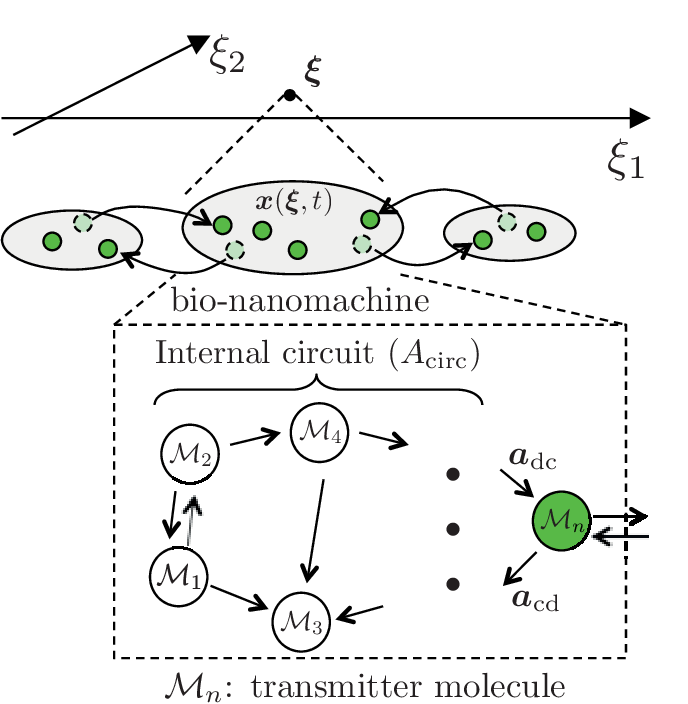}
\caption{Molecular communication network. 
The chemical concentrations inside cells are approximately modeled by
 the continuous gradient of concentrations ${\bm x}({\bm \xi}, t)$.
}
\label{sysdesc-fig}
\end{figure}

We consider molecular communication networks in a two dimensional 
 medium $\Omega:=\Xi_1 \times \Xi_2 = [0,L_{1}] \times [0,L_{2}]$ where a large
number of cells communicate using a single diffusible
transmitter molecule as illustrated in Fig. \ref{sysdesc-fig}. 
Each cell contains a genetically identical gene regulatory circuit that serves as a
functional module such as bistable switch \cite{Gardner2000}, oscillator
\cite{Elowitz2000}, logic gates \cite{Moon2012}, and so on.  
The circuit is composed of $n\!-\!1$ molecular species
$\mathcal{M}_1, \mathcal{M}_2, \cdots, \mathcal{M}_{n-1}$ and is 
called internal circuit (see Fig. \ref{sysdesc-fig}). 
The molecules $\mathcal{M}_1, \mathcal{M}_2, \cdots,
\mathcal{M}_{n-1}$ are, for example, transcription factors and
associated mRNAs.
In order for the cells to communicate, cells produce a transmitter
molecule $\mathcal{M}_n$ and release it to the medium.
An example of such transmitter molecule is N-Acyl homoserine lactone (AHL),
which is a small molecule that can diffuse through cell membrane and is used
for quorum sensing of gram-negative bacteria. %
It should be noted that, in engineering applications, one could use engineered bio-nanomachines such as vesicles instead of living cells. %

\par
\smallskip
When the cells are densely populated in the medium, the chemical
levels in the cells can be approximately modeled as the continuous gradient of
the molecular concentrations inside cells.
Let $x_i({\bm \xi}, t)$ denote the concentration of
$\mathcal{M}_i~(i=1,2,\cdots,n)$ at position ${\bm \xi}:=[\xi_1, \xi_2]^T \in \Omega$ and at time
$t$, and define ${\bm x}({\bm \xi}, t) := [x_1({\bm \xi},t), x_2({\bm \xi},t), \cdots,
x_n({\bm \xi},t)]^T$.
The dynamics of the concentrations of $\mathcal{M}_1, \mathcal{M}_2,
\cdots, \mathcal{M}_n$ are then modeled by 
\begin{align}
\frac{\partial {\bm x}({\bm \xi}, t)}{\partial t} = f({\bm x}({\bm \xi}, t)) + D
\left( \frac{\partial^2 {\bm x}({\bm \xi}, t)}{\partial \xi_1^2} + \frac{\partial^2
 {\bm x}({\bm \xi}, t)}{\partial \xi_2^2} \right),
\label{nonlinear-eq}
\end{align}
where $f(\cdot)$ is a function %
representing the
rate of production and degradation of the molecules.
The matrix $D$ is defined as $D := \mathrm{diag}(0,\cdots,0,\mu) \in \mathbb{R}_{0+}^{n
\times n}$ with a diffusion coefficient $\mu$. 
It should be noted that only the $(n,n)$-th entry is non-zero, since 
only the transmitter molecule $\mathcal{M}_n$ can diffuse between cells.
We consider the Neumann boundary condition 
\begin{align}
\vec{\bm n}\cdot\nabla {\bm x}({\bm \xi}, t) = 0;~~\forall{\bm \xi} \in
 \partial \Omega, 
\end{align}
where $\vec{\bm n}$ is a vector normal to the boundary 
$\partial \Omega$.
Throughout this paper, we assume that the model (\ref{nonlinear-eq}), the boundary condition
and initial values are well-posed (see \cite{Morgan1989} and Chapter 14 of \cite{Smoller1994})
and that the Jacobian of $f(\cdot)$ is well-defined at ${\bm x}$
satisfying $f({\bm x}) = 0$.

\par 
\smallskip 
Let $\bar{{\bm x}} \in \mathbb{R}_{0+}^{n}$ denote an equilibrium
state of a single cell, {\it i.e.}, $f(\bar{\bm x}) = 0$. 
It follows that the right-hand side of (\ref{nonlinear-eq}) becomes zero when ${\bm x}({\bm \xi},t) =
\bar{{\bm x}}$ for all ${\bm \xi} \in \Omega$, because the concentration
of each molecular species is uniform over the space and the spatial derivatives in
(\ref{nonlinear-eq}) become zero. 
This implies that $\bar{\bm x}$ is a spatially homogeneous, or spatially
uniform, equilibrium state of the system (\ref{nonlinear-eq}). 
It should be noted that the spatially homogeneous equilibrium state
might not be unique.
The local dynamics around $\bar{\bm x}$ are modeled by the following linearized
model of (\ref{nonlinear-eq}).
\begin{align}
\frac{\partial \tilde{\bm x}({\bm \xi}, t)}{\partial t} = A \tilde{\bm
 {x}}({\bm \xi}, t) + D
\left( \frac{\partial^2 \tilde{{\bm x}}({\bm \xi}, t)}{\partial \xi_1^2} + \frac{\partial^2
 \tilde{{\bm x}}({\bm \xi}, t)}{\partial \xi_2^2} \right), 
\label{linear-eq}
\end{align}
where $A \in \mathbb{R}^{n \times n}$ is the Jacobian of $f(\cdot)$
 at $\bar{{\bm x}}$, and $\tilde{\bm x}({\bm \xi}, t)$ %
is defined by $\tilde{{\bm x}}({\bm \xi}, t) := {\bm x}({\bm \xi}, t) - \bar{{\bm
x}}$. Note that the system (\ref{linear-eq}) is an infinite dimensional linear
 time-invariant system.

\par
\smallskip

 \par
 \smallskip
 In Section \ref{conttheo-sec}, we provide a systematic analysis tool for
 studying stability of the equilibrium point ${\bar {\bm x}}$ using the
 linearized model (\ref{linear-eq}).
We then show that cells can form spatially inhomogeneous gradient
 of concentrations over the population when the system (\ref{linear-eq}) is locally
 unstable and satisfies certain conditions.

\subsection{Control theoretic formulation of biomolecular communication networks}

In this section, we introduce a control theoretic model representation
 of the biomolecular communication network system. 
We show that the infinite dimensional system (\ref{linear-eq}) can be 
decomposed into an infinite number of finite dimensional single-input single-output (SISO)
subsystems that account for the dynamics of individual spatial modes,
which will be a key result in developing a systematic stability analysis
tool in the next section. 

\par
\begin{figure}[tb]
\centering
\includegraphics[clip,width=8.5cm]{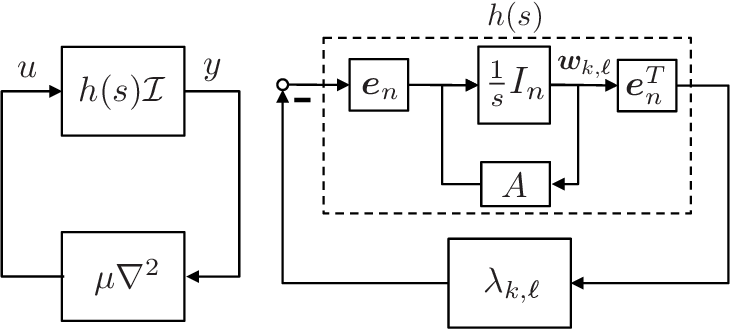}
\caption{(Left) Feedback system representation of biomolecular
 communication network. (Right) Decomposed subsystem $\Sigma_{k,\ell}$.}
\label{block-fig}
\end{figure}

\smallskip
The linearized model (\ref{linear-eq}) can be expressed as 
\begin{align}
&\displaystyle \frac{\partial {\tilde{{\bm x}}}({\bm \xi}, t)}{\partial t}=A \tilde{{\bm
 x}}({\bm \xi}, t) + {\bm e}_n u({\bm \xi}, t) 
\label{statespace-eq1}
\\
&y({\bm \xi}, t) = {\bm e}_n^T \tilde{{\bm x}}({\bm \xi}, t),
\label{statespace-eq2}
\end{align}
where $u({\bm \xi}, t) := \mu \nabla^2 y({\bm \xi}, t)$ with the
Laplace operator
\begin{align}
\nabla^2 := \frac{\partial^2}{\partial \xi_1^2} + \frac{\partial^2}{\partial \xi_2^2}
\end{align}
 and ${\bm e}_n := [0,\cdots, 0,1]^T \in
\mathbb{R}^{n}$. 
The equations (\ref{statespace-eq1}) and (\ref{statespace-eq2}) imply that for each fixed position ${\bm
\xi}$, the dynamics of the local reactions inside a single cell can be
modeled as a SISO system with the input $u({\bm \xi}, t)$ and the output
$y({\bm \xi}, t)$. Note that $y({\bm \xi}, t)$ is the concentration of
the transmitter molecule $\mathcal{M}_n$ at position ${\bm \xi}$.
For each ${\bm \xi}$, the transfer function from $u$ to $y$ is obtained as 
\begin{align}
h(s) := {\bm e}_n^T (sI_n - A)^{-1} {\bm e}_n~ (=:n(s)/d(s)),
\label{h-def}
\end{align}
 where we define $n(s)$ and $d(s)$ as the numerator and the denominator 
of $h(s)$, respectively.
As a result, the system can be illustrated as the feedback system shown in
Fig. \ref{block-fig} (Left), 
 where $\mathcal{I}$ is an identity operator. %
Note that the feedback system can be viewed as a 
multi-agent dynamical system, where the homogeneous agents $h(s)$ communicate with the
nearest neighbors by the Laplace operator $\mu \nabla^2$. 
The stability analysis methods of a finite dimensional version of such systems
were established in \cite{Hara2009, Hara2014, Nakamura2014}. 
Here we use the same approach to analyzing the stability of the infinite
dimensional system (\ref{linear-eq}).

\medskip
\noindent
{\bf Assumption 1.}~{\it We assume $(A, {\bm e}_n)$ is stabilizable, 
$(A,{\bm e_n^T})$ is detectable, and $h(s)$ does not have zeros on the imaginary axis, that is, 
$n(j\omega) \neq 0$ for all $\omega \in \mathbb{R}$. }

\medskip
\noindent
Stabilizability and detectability guarantees the stability of the 
internal states, or molecular concentrations, that one cannot aware of
and/or influence, if any. The readers are referred to Chapters 14 and 16
of \cite{Hespanha2009} for the mathematical details.
We also note that the non-existence of the zeros of $h(s)$ on the
imaginary axis is not restrictive in many biological applications as
shown after Lemma 1 in Section \ref{conttheo-sec}.

\par
\smallskip
The stability analysis of the system
(\ref{linear-eq}) is not necessarily straightforward due to the infinite
dimensionality of the system. 
In what follows, we show that the infinite dimensional system can be
decomposed into an infinite number of mathematically
tractable finite dimensional subsystems.
The Laplace operator $\nabla^2$ can be written as 
\begin{align}
\nabla^2 = \mathcal{I}_{\Xi_{2}}\otimes \frac{\partial^2}{\partial \xi_1^2} +
\frac{\partial^2}{\partial \xi_2^2} \otimes \mathcal{I}_{\Xi_1}
\label{nabla2-def}
\end{align}
with the tensor product $\otimes$ of the operators and the identity
operators $\mathcal{I}_{\Xi_i}$ on $\Xi_{i} = [0,L_i]~(i=1,2)$.
The eigenvalues $\lambda_{k,\ell}$ and the associated eigenfunctions 
$\phi_{k,\ell}({\bm \xi})$ of $- \mu \nabla^2$ are 
\begin{align}
&\lambda_{k,\ell} := \mu \left( \left(\frac{k \pi}{L_1} \right)^2 
 + \left(\frac{\ell \pi}{L_2} \right)^2 \right) \label{lambda-def}\\
&\phi_{k,\ell}({\bm \xi}) := \cos \left(\frac{k \pi \xi_1}{L_1} \right)
  \cos \left(\frac{\ell \pi \xi_2}{L_2} \right)
\label{phikl-def}
\end{align}
with $k=0,1,2,\cdots$ and $\ell=0,1,2,\cdots$.
This implies that we can diagonalize the Laplace operator by 
$(\mathcal{F} \otimes \mathcal{F}) \nabla^2 (\mathcal{F}^{-1}\otimes
\mathcal{F}^{-1})$
with the Fourier transform operator $\mathcal{F}$. 
Since $(\mathcal{F}^{-1} \otimes \mathcal{F}^{-1})(\mathcal{I}_{\Xi_2}
\otimes h(s) \mathcal{I}_{\Xi_1} + h(s) \mathcal{I}_{\Xi_2} \otimes \mathcal{I}_{\Xi_1}) (\mathcal{F} \otimes \mathcal{F})=\mathcal{I}_{\Xi_2}
\otimes h(s) \mathcal{I}_{\Xi_1}+ h(s) \mathcal{I}_{\Xi_2} \otimes \mathcal{I}_{\Xi_1}$,
the feedback system in Fig. \ref{block-fig} (Left) can be decomposed
into an infinite number of subsystems $\Sigma_{k,\ell}~(k=0,1,2,\cdots, \ell=0,1,2,\cdots)$ depicted in
Fig. \ref{block-fig} (Right).
The readers are referred to, for example, Chapters 4 and 10 of \cite{Strauss2008} for the detailed derivation
of the eigenvalues and the eigenfunctions.

\par
\smallskip
This decomposition corresponds to the spatial modal
decomposition of the molecular communication network system, which 
is essentially the method of the separation of variables (see
\cite{Holland1978} and Chapter 4 of \cite{Strauss2008}).
Thus, the dynamics of each subsystem $\Sigma_{k,\ell}$ 
represent those of each spatial mode $\phi_{k,\ell}({\bm \xi})$, and  it follows that 
\begin{align}
\tilde{{\bm x}}({\bm \xi}, t) = \sum_{k=0}^{\infty}\sum_{\ell=0}^{\infty} 
{\bm w}_{k,\ell}(t)
\phi_{k,\ell}({\bm \xi}),
\label{tildex-eq}
\end{align}
where ${\bm w}_{k,\ell}$ is the state of the subsystem
$\Sigma_{k,\ell}$. %

\par
\smallskip
This intuitively suggests that the concentrations perturbed around the homogeneous
equilibrium $\bar{\bm x}$ eventually come back to $\bar{\bm x}$, when
the growth rates of all the spatial modes are negative. 
On the other hand, if the growth rate of a non-zero spatial mode,
$\phi_{k,\ell}(\xi)$ with $k+\ell \ge 1$, is positive, we expect the
formation of a spatially periodic pattern of 
concentration gradient, or a Turing pattern.
Therefore, the stability analysis of the molecular communication
network is reduced to checking the stability of the finite-order subsystems
$\Sigma_{k,\ell}$. 

\medskip
\noindent
{\bf Remark 1.~}
In general, the eigenvalues and the eigenfunctions defined in
(\ref{lambda-def}) and (\ref{phikl-def}) can be written
as 
\begin{align}
\lambda_{{\bm k}} = \mu \sum_{i=1}^{N}
\left(\frac{\lambda_{k_i}\pi}{L_i}\right)^2, 
\phi_{{\bm k}}({\bm \xi}) = \prod_{i=1}^{N}\cos
\left(
\frac{k_i \pi \xi_i}{L_i}
\right) \label{Ndimeig-eq}
\end{align}
when the spatial domain is $N$ dimension and the Neumann boundary condition
is imposed, where ${\bm k}$ is the set of subindices $(k_1, k_2, \cdots,
k_N)$ concerning the spatial dimension and $L_i$ is the length of the $i$-th dimension.
The pair of eigenvalues and eigenfunctions depends on the boundary condition.
For example, the eigenvalues for the Dirichlet boundary condition are given by
(\ref{Ndimeig-eq}) with $\sum_{i=1}^{N}k_i \ge 1$. In other words, the zero
eigenvalue $\lambda_{0,0,\cdots,0} = 0$ is not an eigenvalue for the Dirichlet condition.
The associated eigenfunctions are $\phi_{{\bm k}}({\bm \xi}) = \prod_{i=1}^{N}\sin \left(k_i \pi \xi_i/L_i\right)$.
We refer the readers to Chapters 10 and 11 of \cite{Strauss2008} for
other boundary conditions. 
In the rest of this paper, we focus on the case of two dimensional
spatial domain $\Omega$ with the Neumann boundary condition, since biomolecular communication
networks would be implemented on a two dimensional surface.

\section{Stability analysis method for linearized model}
\label{conttheo-sec}
In this section, we provide a systematic local stability analysis tool
for the linearized model (\ref{linear-eq}) of the molecular
communication networks using a control theoretic tool. 
The method is demonstrated on a novel activator-repressor-diffuser
system. In particular, we show that the instability counterpart of the
method is useful for the design and analysis of coordinated spatial
pattern formation of the synthetic biocircuits.
 
\subsection{Graphical stability analysis method based on root locus}

\par
\smallskip
Suppose the internal circuit inside a cell is composed of
$n \ge 2$ molecular species, and let the matrix $A \in \mathbb{R}^{n \times n}$ be partitioned as  
\begin{equation}
A = \left[
\begin{array}{c:c}
{A}_{{\rm circ}} & {\bm a}_{{\rm cd}} \\ \hdashline
{\bm a}_{{\rm dc}} & a_{{\rm diff}}
\end{array}
\right] 
\end{equation}
with $A_{{\rm circ}} \in \mathbb{R}^{(n-1) \times (n-1)}, {\bm a}_{{\rm
cd}} \in \mathbb{R}^{n-1}, {\bm a}_{{\rm dc}} \in \mathbb{R}^{1 \times
(n-1)}$ and $a_{{\rm diff}} < 0$.
Then, the matrix $A_{\rm circ}$ represents the structure of the internal
circuit, and $a_{{\rm diff}} (< 0)$ is the degradation rate of 
the single diffusible molecule. The vectors ${\bm a}_{{\rm cd}}$ and
${\bm a}_{\rm dc}$ correspond to the interactions between the internal
circuit molecules $\mathcal{M}_1, \mathcal{M}_2, \cdots,
\mathcal{M}_{n-1}$ and the diffuser $\mathcal{M}_n$ (see Fig. \ref{sysdesc-fig}).
When $n=1$, we define $A = a_{\rm diff}$. 

\par
\smallskip
Let a polynomial $p_{\lambda}(s)$ be defined by
\begin{align}
p_{\lambda}(s) := d(s) + \lambda n(s)
\end{align}
with $d(s)$ and $n(s)$ in (\ref{h-def}). 
The characteristic polynomial of each subsystem $\Sigma_{k,\ell}$ is then
 written as $p_{\lambda_{k,\ell}}(s) = d(s) + \lambda_{k,\ell}n(s)$. 
The following lemma provides a physical interpretation of the
poles and zeros of $h(s)$ using $A$ and $A_{\rm circ}$.

\medskip
\noindent
{\it 
{\bf Lemma 1.~}
Consider the molecular communication network system (\ref{linear-eq}). 
Then, the characteristic polynomial $p_\lambda(s)$ of the system can be written as 
\begin{align}
p_\lambda(s)\!=\!d(s) + \lambda n(s)\!=\!|sI_n - A| + \lambda |sI_{n-1} -
 A_{\rm circ}|, 
\label{prop2-eq}
\end{align}
where we define $|sI_0 - A_{\rm circ}| = 1$ when $n=1$.
}%

\par
\medskip
The proof can be found in Appendix. %
The lemma states that the zeros of $h(s)$ are determined from the 
dynamics of the internal circuit specified by $A_{\rm circ}$, while the poles are determined from the overall reaction
dynamics including the diffuser $\mathcal{M}_n$. 
Regarding Assumption 1, we note that $n(s)=|sI_{n-1}-A_{\rm circ}| = 0$ does not have roots on the
imaginary axis for almost all $A_{\rm circ}$ when all diagonal entries
of $A_{\rm circ}$ are non-zero.  
This is because the coefficient of the $s^{n-2}$ term is given by ${\rm
Tr}(A_{\rm circ})$. 
In fact, the diagonal entries of $A_{\rm circ}$ are non-zero in many
biomolecular systems due to degradation and autoregulation.

\par
\smallskip
It follows that the subsystem $\Sigma_{k,\ell}$ is asymptotically stable, if and
 only if all the roots of $p_{\lambda_{k,\ell}}(s) = 0$ lie in the open left-half
 complex plane, also called Hurwitz.
Since each subsystem represents the dynamics of each spatial mode as
shown in the previous section, the stability of the molecular
communication network (\ref{linear-eq}) is guaranteed, 
if and only if all of the infinite number of finite dimensional subsystems
$\Sigma_{k,\ell}~(k=0,1,2,\cdots, \ell=0,1,2,\cdots)$ are stable. 

\medskip
\noindent
{\bf Proposition 1.~}
{\it Suppose Assumption 1 holds. The molecular communication network (\ref{linear-eq}) is exponentially stable 
if and only if $p_{\lambda_{k,\ell}}(s)$ is Hurwitz for all $k, \ell \in
\mathbb{N}_0$. 
}

\medskip
\noindent
{\bf Proof.}
We show the proof for the case of one-dimensional spatial domain
 in order to avoid notational clutter. However, generalization to
multiple spatial dimensions is straightforward.
Let $\{\sigma_{k,i}\}_{i=1}^{n}$ denote the roots of $p_{\lambda_k}(s) = 0$.
Then, it follows that 
\begin{align}
{\bm w}_{k}(t) =\sum_{i=1}^{n} c_{k,i} e^{\sigma_{k,i} t} {\bm
 w}_{k}(0), 
\label{wk-eq}
\end{align}
where $c_{k,i}~(i=1,2,\cdots,n)$ are some constant. 

\par
\smallskip
($\Rightarrow$) Suppose all of the polynomials $p_{\lambda_k}(s) = 0~(k\in\mathbb{N}_0)$
are Hurwitz. This means that there exists a positive real number
$\epsilon_1$ such that $\mathrm{Re}[\sigma_{k,i}] < -\epsilon_1 (< 0)$ for all $k\in\mathbb{N}_0$.
In the limit of $k \rightarrow \infty$, the eigenvalue $\lambda_k
\rightarrow \infty$, and $n-1$ roots of $p_{\lambda_{k}}(s)$
converge to those of $n(s)=0$ and one root converges to $-\infty$, known 
as Butterworth pattern \cite{Levine1996}, since $n(s)$ is the $n -
1$-th order polynomial as shown in Lemma 1. Note that this rules out the
possibility that the root asymptotically approaches to the imaginary
axis.  Let the roots of $n(s)=0$ be 
$\{\sigma_{\infty,i}\}_{i=1}^{{n-1}}$.
Since $n(s) = 0$ does not have roots on the imaginary axis from
Assumption 1, there exists a positive real number $\epsilon_2$ such that 
$\mathrm{Re}[\sigma_{\infty,i}] < -\epsilon_2 (< 0)$ for all
$i=1,2,\cdots,n-1$. 
Substituting (\ref{wk-eq}) into (\ref{tildex-eq}) and taking the
integral over the space $\Omega$ leads to 
\begin{align}
\int_\Omega \|\tilde{{\bm x}}(\xi, t)\| d\xi\!=\!& \int_\Omega 
\left\|\sum_{k=0}^{\infty}
\left(\sum_{i=1}^{n} c_{k,i} e^{\sigma_{k,i} t} {\bm
 w}_{k}(0) \right) 
\phi_{k}(\xi)\right\| d\xi \notag \\
\le &
C e^{-\epsilon t} \int_\Omega \left\| \sum_{k=0}^{\infty}{\bm w}_{k}(0)
 \phi_{k}(\xi)\right\| d \xi \notag 
\\ = &C e^{-\epsilon t} \int_\Omega \left\| \tilde{\bm x}(\xi, 0)\right\| d
 \xi, \label{int-eq}
\end{align}
where $C$ is some constant and $\epsilon := \min(\epsilon_1, \epsilon_2)$. 
Note that $\phi_k(\xi) = \cos(k\pi\xi/L_1)$ and the summation over $\ell$ in (\ref{tildex-eq}) 
is dropped due to one dimensional spatial domain.

\par
\smallskip
($\Leftarrow$) Suppose one of the polynomials $p_{\lambda_k}(s) = 0$ is
not Hurwitz, which implies the existence of $k_0 \in \mathbb{N}_0$ and $i_0 \in
\{1,2,\cdots,n\}$ such that $\mathrm{Re}[\sigma_{k_0,i_0}] \ge 0$. 
This implies that the growth rate of the mode $\phi_{k_0}(\xi)$ is
non-negative, and the right-hand side of the first equality of 
(\ref{int-eq}) is bounded from below by
$Ce^{\mathrm{Re}[\sigma_{k_0,i_0}]t}\|{\bm w}_{k_0}(0)\|$ with some constant $C$.
$\hfill\Box$

\medskip
Proposition 1 allows us to analyze the stability of the infinite dimensional
system (\ref{linear-eq}) using the finite dimensional subsystems
$\Sigma_{k,\ell}~(k=0,1,2,\cdots, \ell=0,1,2,\cdots)$. 
It should be noted that there are still infinite subsystems
$\Sigma_{k,\ell}~(k=0,1,2,\cdots, \ell=0,1,2,\cdots)$, although each
subsystem is a finite dimensional system.
In what follows, we show that the root locus method \cite{Levine1996}
allows us to systematically analyze the stability of the infinite number
of subsystems $\Sigma_{k,\ell}$.

\par
\smallskip
It follows that the roots of 
$p_{\lambda_{k,\ell}}(s) = 0 ~(k=0,1,2,\cdots,\ell=0,1,2,\cdots)$ 
lie on the root locus of the polynomial $p_{\lambda}(s) = d(s) + \lambda
n(s)$, which is the characteristic polynomial of a feedback system
composed of $h(s)$ and a feedback gain $\lambda$ (see
Fig. \ref{block-fig} (Right)).
Moreover, the definition (\ref{lambda-def}) implies that the feedback gain 
$\lambda = \lambda_{k,\ell}$ monotonically increases in terms of $k$ and $\ell$.
Thus, the root locus starts from the point where $\lambda =
\lambda_{0,0} = 0$ and converges to the point where $\lambda =
\lim_{k,\ell\rightarrow \infty} \lambda_{k,\ell}$. %
This observation leads to the following graphical algorithm for
stability analysis.

\medskip
\noindent
{\bf Algorithm 1.~}
{\it Given the dynamics of a single cell $h(s)$, we draw the root
locus $\mathcal{R}$ of a negative feedback system composed of $h(s)$ and a constant
feedback gain $\lambda$ with varying $\lambda \in [0,\infty]$. If the
root locus $\mathcal{R}$ lies inside the open left-half complex plane
$\mathbb{C}_{-}$, the molecular communication network (\ref{linear-eq})
is stable.
}

\medskip
\noindent
This algorithm allows us to check the local stability of the molecular
communication network system (\ref{linear-eq}) using the transfer function of
a single cell $h(s)$ and its root locus. 
Thus, the stability analysis of the infinite dimensional system is 
greatly simplified. 

\medskip
\noindent
{\bf Remark 2.~}
The reason that the root locus gives only a sufficient condition for
stability is
because the root locus is continuous in terms of the feedback
gain $\lambda$, while the poles of the subsystems $\Sigma_{k,\ell}~(k=0,1,2,\cdots, \ell = 0,1,2,\cdots)$ 
are discrete. 
In practice, however, the length of the spatial domain $L_1$ and $L_2$ is
sufficiently large such that the discrete feedback gains
$\lambda_{k,\ell} = \mu\left((k\pi/L_1)^2 + (\ell \pi/L_2)^2
\right)~(k=0,1,2,\cdots, \ell=0,1,2,\cdots)$ are close to each other as
shown in Section \ref{ard-sec}, 
and the above algorithm can be practically considered as necessary and
sufficient. 

\par
\medskip
It is worth noting that the root locus can be systematically drawn based
on a set of known rules \cite{Levine1996}. 
Specifically, the locus $\mathcal{R}$ starts from the $n$ poles of 
$h(s)$, and $n-1$ roots converge to the zeros of $h(s)$ and one pole
goes to $+\infty$ or $-\infty$, since the relative degree of $h(s)$ is
one.

\medskip
\noindent
{\bf Lemma 2.~}
{\it 
The root locus $\mathcal{R}$ starts from $\mathrm{spec}(A)$, 
and $n-1$ roots converge to $\mathrm{spec}(A_{\rm circ})$. %
}

\medskip
\noindent
This is a direct consequence of Lemma 1 and the properties of the root
locus \cite{Levine1996}. 
This lemma implies that the start and the terminal point of the root locus
$\mathcal{R}$ can be characterized by the dynamics of the entire circuit
and the internal circuit, respectively. 
This intuition is helpful from a synthetic biology viewpoint to 
determine the structure of the internal biocircuit to guarantee the
stability/instability of the system.

\medskip
\noindent
{\bf Remark~3.}
Arcak \cite{Arcak2011} showed algebraic conditions to guarantee  
the convergence of reaction-diffusion systems to spatially uniform
solutions. 
When applied to our problem, Theorem 1 of \cite{Arcak2011} provides a
sufficient condition for all the subsystems $\Sigma_{k,\ell}$ with
$k+\ell \ge 1$ to be exponentially stable.
Thus, the condition is sufficient for the stability of the spatially
homogeneous equilibrium $\bar{{\bm x}}$, provided that $A$
is Hurwitz, or equivalently $\Sigma_{0,0}$ is stable.
However, this sufficient condition is conservative compared to 
Proposition 1, which is necessary and sufficient, and Algorithm 1 of this paper, since it is based on constituting
a common Lyapunov solution that verifies the stability of all of the subsystems $\Sigma_{k,\ell}$. 
On the other hand, theorems shown in \cite{Arcak2011} can be used to 
certify the spatially uniformity of non-stationary solutions such as 
oscillations, while Proposition 1 cannot.

\subsection{Example: activator-repressor-diffuser system}
\label{ard-sec}
\begin{figure}[tb]
\centering
\includegraphics[clip,width=8cm]{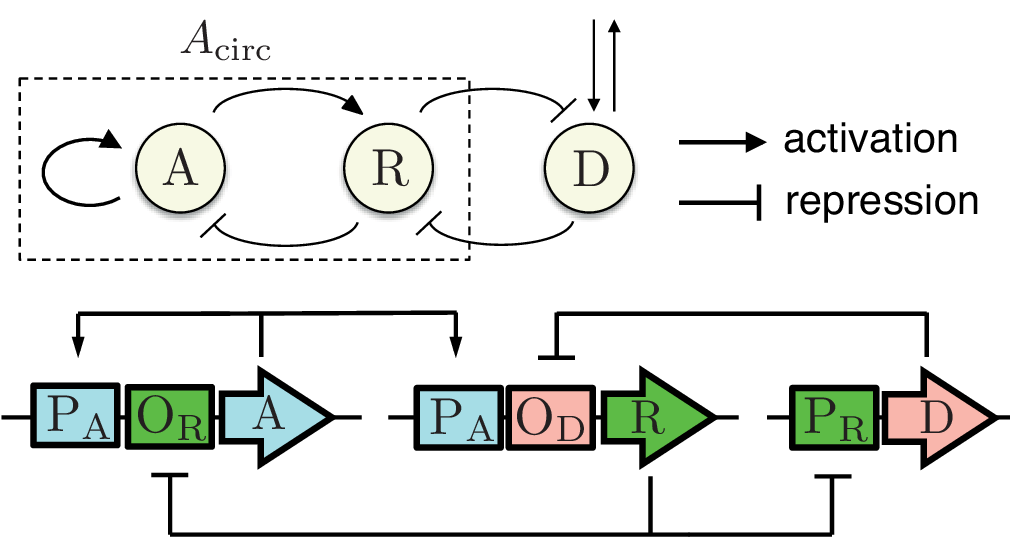}
\caption{Schematic diagram of activator-repressor-diffuser system}
\label{ard-syst-fig}
\end{figure}

We illustrate the graphical stability analysis method using a 
hypothetical biomolecular network shown in Fig. \ref{ard-syst-fig}. The system
is composed of activator-repressor internal circuit and a single
diffuser molecule that allows cell-to-cell communication. 
In Fig. \ref{ard-syst-fig}, $A, R$ and $D$ stand for activator, 
repressor and diffuser, respectively.
In the context of synthetic biocircuit design, one can choose 
an AraC protein as an (inducible) activator, {\it i.e.,} $A$ in Fig. \ref{ard-syst-fig}, and the araBAD operon
as the corresponding promoter ${\rm P}_A$.
For the repressor $R$, commonly used TetR and LacI proteins and the
corresponding promoters can be used.
The diffuser is implemented with AHL, which is a bacterial quorum
sensing molecule. 

\par
\smallskip
Let $a({\bm \xi}, t), r({\bm \xi}, t)$ and $d({\bm \xi}, t)$ denote the concentration of
activator, repressor and diffuser, respectively. The dynamics of the
molecular communication network system are then modeled by 
\begin{align}
&\dot{a} = -\delta_a a + \gamma_a \frac{{a}^{2}}{K_a^2 + a^{2}}
 \frac{K_r^2}{K_r^2 + r^{2}} \label{a-eq}\\
 &\dot{r} = -\delta_r r + \gamma_r \frac{{a}^{2}}{K_a^2 + a^{2}}
  \frac{K_d^2}{K_d^2 + d^{2}}\\
 &\dot{d} = -\delta_d d + \gamma_d\frac{K_r^2}{K_r^2 + r^{2}} + \mu \nabla^2 d, 
\label{d-eq}
\end{align}
where $\delta_*~(*=a,r,d)$ and $\gamma_*~(*=a,r,d)$ denote the
 degradation rates and the production rates of each molecular species, respectively.

\par
\smallskip
\noindent
{\bf Case A. (Stable case): }
Suppose the half-lives of activator, repressor and diffuser are 19.8
minutes, 55.5 minutes and 138.6 minutes, respectively, which equates to 
$\delta_a = 0.035 {\rm min}^{-1}$, $\delta_r = 0.0125 {\rm min}^{-1}$ and
$\delta_d = 0.0050 {\rm min}^{-1}$. 
The production rates of each molecular species are $\gamma_a = 2.5 \mu{\rm
M}\cdot {\rm min}^{-1}$, $\gamma_r = 1.65 \mu{\rm
M}\cdot {\rm min}^{-1}$, $\gamma_d = 0.25 \mu{\rm
M}\cdot {\rm min}^{-1}$, and the Michaelis-Menten constants are $K_a =
K_r = K_d = 10 \mu{\rm M}$. The diffusion rate is $\mu = 3.0 \times 10^{-4}$
mm$^2 \cdot {\rm min}^{-1}$. 
These parameter values are consistent with widely used parameters for
numerical simulations in synthetic biology up to the order of magnitude (see \cite{Basu2005} for 
example). We leave the discussion on the realizability of such biomolecular
communication networks in Remark 4 and focus on the numerical illustration of the root locus based analysis
method here.

\par
\smallskip
Substituting the parameters into the model (\ref{a-eq})--(\ref{d-eq}), we 
first calculate the spatially homogeneous equilibrium point of the system
as $[a_*, r_*, d_*]^T = [9.33, 16.0, 16.8]^T$. 
The Jacobian matrix $A$ around this equilibrium point is 
\begin{align}
A = 
10^{-2} \times \left[
\begin{array}{cc:c}
0.243 & -2.93 & 0\\
2.29 & -1.25 & -1.76 \\ \hdashline
0 & -0.757 & -0.500
\end{array}
\right]
\end{align}
The transfer function of a single cell $h(s)$ is then calculated from (\ref{prop2-eq}) as 
\begin{align}
h(s) = \frac{s^2 + 1.01 \!\times \!10^{-2} s + 6.43 \!\times \! 10^{-4}}
{s^3 + 1.51 \!\times \! 10^{-2} s^2 + 5.60 \!\times \! 10^{-4} s +
 3.54\! \times \!10^{-6}}. \notag
\end{align}

\par
\begin{figure}[tb]
\centering
\includegraphics[clip, width=9cm]{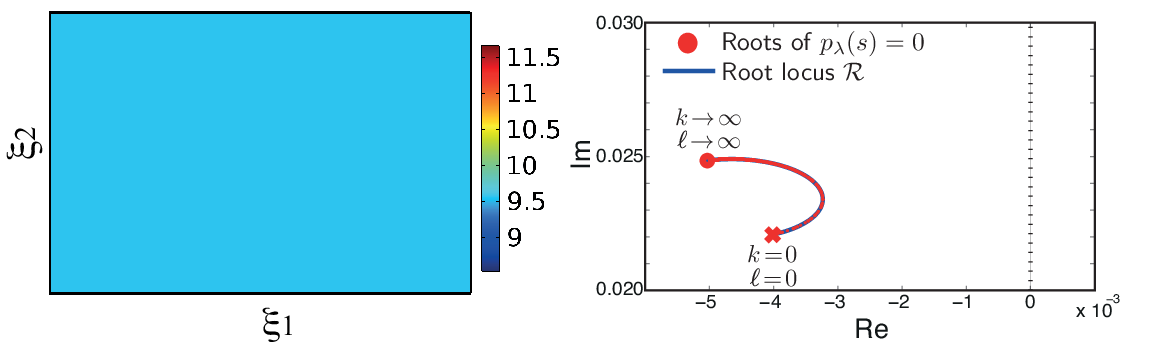}
\caption{Case A: (Left) Concentration gradient of the activator $A$ over
 cell population at $t=2855$ min. (Right) Root locus
 $\mathcal{R}$ of the single cell dynamics $h(s)$.}
\label{nopattern-eg-fig}
\end{figure}

\smallskip
Figure \ref{nopattern-eg-fig} (Right) illustrates the root locus
$\mathcal{R}$ drawn by Algorithm 1. 
The eigenvalues of $A$ and $A_{\rm circ}$ are 
$\mathrm{spec}(A) = \{-0.00402 \pm 0.0221j, -0.00702 \}$
and 
$\mathrm{spec}(A_{\rm circ}) = \{-0.00503 \pm 0.0248j\}$, and we can see
that these eigenvalues correspond to the initial and the terminal points of
the root locus $\mathcal{R}$. 
Since the root locus $\mathcal{R}$ stays in the left-half complex plane,
the equilibrium point of the molecular communication network is locally stable.

\par
\smallskip
Figure \ref{nopattern-eg-fig} (Left) is the concentration gradient of
the activator $a(\xi,t)$ at $t = 2855$ min. As expected from
the graphical stability analysis result, the concentration converged to 
the spatially homogeneous equilibrium point. In the simulation, we used
$L_1 = 3 {\rm mm}$  and $L_2 = 2 {\rm mm}$, and the initial values $a({\bm \xi}, 0), r({\bm \xi}, 0)$ and $d({\bm
\xi}, 0)$ were set as $x_* (1 + 0.01 \sum_{k=1}^{5}\sum_{\ell=1}^{5} \psi_{k,\ell}(\xi_1,\xi_2))$, 
where $x_*$ is the equilibrium point of $a({\bm \xi}, 0), r({\bm \xi}, 0)$ and $d({\bm
\xi}, 0)$, respectively, and 
\begin{align}
{\small
\psi_{k,\ell}(\xi_1, \xi_2)\!:=\!\cos\left(
\frac{k\pi}{L_1} \left(
\xi_1\!-\!\frac{L_1}{2}
\right)
\right)
\!
\cos\left(
\frac{\ell\pi}{L_2}
\left(
\xi_2\!-\!\frac{L_2}{2}
\right)
\right).
}\notag
\end{align}
The full temporal dynamics is available in a movie format
\cite{Hori2015Tech}, where the interval between frames is 5 minutes.
We used COMSOL$\textregistered$ Multiphysics 4.4 for 
the simulation.

\par
\begin{figure}[tb]
\centering
\includegraphics[clip, width=9cm]{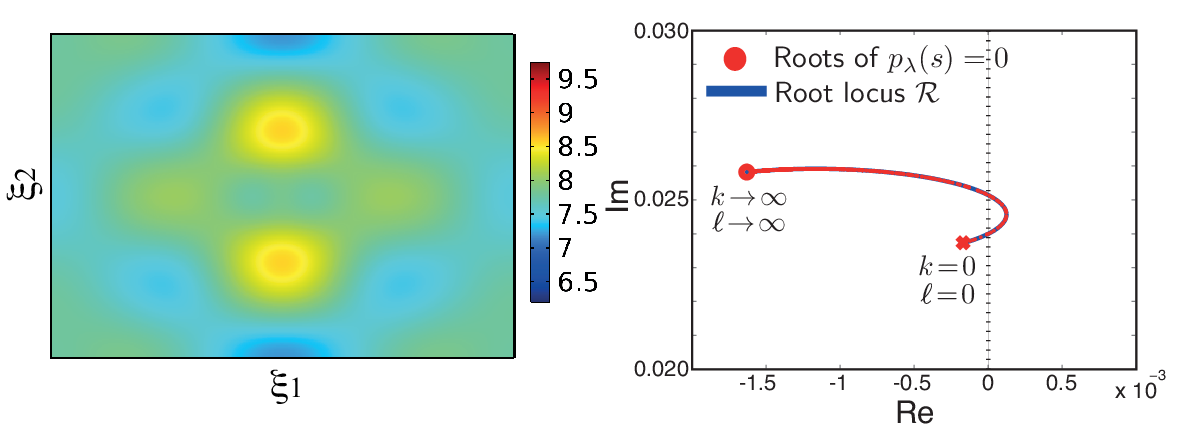}
\caption{Case B: (Left) Concentration gradient of the activator $A$ over
 cell population at $t=2855$ min. (Right) Root locus
 $\mathcal{R}$ of the single cell dynamics $h(s)$.}
\label{pattern-eg-fig}
\end{figure}

\smallskip
\noindent
{\bf Case B. (Unstable case): }
Let $\gamma_d = 0.30 \mu{\rm M}\cdot{\rm min^{-1}}$ and the rest of the parameters being kept the same as
Case A. The equilibrium point is then 
$[a_*, r_*, d_*]^T = [7.63, 15.6, 14.5]^T$.
The eigenvalues of $A$ and $A_{\rm circ}$ are calculated as
$
\mathrm{spec}(A) = \{
-0.000169 \pm 0.0237j, -0.00792
\}
$
and 
$\mathrm{spec}(A_{\rm circ}) = \{-0.00163 \pm 0.0258j\}$.
As shown in Fig. \ref{pattern-eg-fig} (Right), the root locus lies in the open right-half complex
plane. In fact, multiple roots of $p_{\lambda}(s) = 0$ lie in the right-half complex plane, implying
that the corresponding subsystems, or spatial modes, are unstable. 

\par
\smallskip
Figure \ref{pattern-eg-fig} (Left) illustrates the concentration gradient
of the activator $A$ at $t=2855$ min. 
Since the equilibrium point is not stable, the concentration does not
converge to the spatially homogeneous equilibrium. In fact, 
a coordinated time-varying spatial pattern appeared as illustrated in
Fig. \ref{pattern-eg-fig} (Right), 
implying that the system fell into limit cycle instead of
converging to a spatially inhomogeneous equilibrium point.
The full temporal dynamics is available in a movie format
\cite{Hori2015Tech}, where the interval between frames is 5 minutes.
We used COMSOL$\textregistered $ Multiphysics 4.4 for 
the simulation.

\par
\smallskip
As Turing \cite{Turing1952} pointed out, this coordinated pattern formation
can be explained by the fact that the subsystems corresponding to the non-zero spatial modes are
destabilized. In the next section, we provide more rigorous argument of 
why and when the pattern formation can be expected.

\medskip
\noindent
{\bf Remark~4.}
The libraries of activators \cite{Rhodius2013}, repressors \cite{Stanton2014} and ribosome binding
sites \cite{Mutalik2013} recently became available for the use in synthetic
biology. These libraries help us tune the rate parameters in a
relatively predictable manner and would be useful for the implementation
of the activator-repressor-diffuser biocircuit. 
In fact, it is encouraging that multiple biocircuits were already implemented
and tuned using these parts libraries \cite{Rhodius2013, Stanton2014, Henrike2015}. 
In the context of biomolecular communication network, the use of zinc finger
proteins and small RNAs were also proposed in \cite{Hsia2012CDC}.

\section{Conditions for Turing Pattern Formation}
\label{turing-sec}

\subsection{Finite mode Turing instability}
As seen in the previous example, molecular communication networks can 
exhibit coordinated spatial patterns based on the passive communication
mechanism, or diffusion. This is because non-zero finite spatial modes
are destabilized and the growth rate of ${\bm w}_{k,\ell}$ in
(\ref{tildex-eq}) is positive.

\par
\smallskip
The instability of the non-zero spatial mode caused by diffusion is
called Turing instability named after Alan Turing, who first pointed out 
the mathematical basis of the emergence of inhomogeneous spatial
patterns in reaction-diffusion equations \cite{Turing1952}. 
In what follows, we introduce a formal definition of finite mode Turing
instability for biomolecular communication networks, then we investigate conditions for the Turing instability.

\medskip
\noindent
{\bf Definition 1. (Finite mode Turing instability)~}
{\it 
Suppose Assumption 1 holds. 
A biomolecular communication network (\ref{nonlinear-eq}) is finite mode Turing unstable around an equilibrium $\bar{{\bm x}}$, 
if

\noindent
(i) $p_{\lambda_{0,0}}(s) = p_{0}(s) = d(s) \neq 0${\rm ;} $\forall s \in \mathbb{C}_{0+}$, and \\
(ii) $p_\lambda(\gamma + s) = 0${\rm ;} $\exists \lambda \in
(0,\infty)$ and $\exists s \in \mathbb{C}_{+}$, where $\gamma =
\max(\mathrm{Re}[s_0], 0)$ and $s_0$ is a root of $n(s) = 0$ with the largest real part.

}%

\par
\medskip
The condition (i) implies that the dynamics of each cell $h(s)$ is stable
around an equilibrium point $\bar{\bm x}$ when there is no diffusion. 
The condition (ii) guarantees that the root locus of $h(s)$ goes into
the open right-half complex plane for some feedback gain
$\lambda > 0$ (see Fig. \ref{block-fig} (Right)). 
In particular, the roots of $n(s) = 0$ are the terminal points of the root
locus (Lemma 2), thus the condition (ii) also implies that the right-most roots of
$p_{\lambda}(s) = 0$ are given when $\lambda$ is a non-zero and finite
value. This guarantees that the dominant unstable spatial mode is
finite and the resulting spatial pattern is of a non-zero finite spatial
mode.
We note that finite mode Turing instability is determined by
the properties of the internal circuit $A_{\rm circ}$ and $A$ but not
the diffusion rate $\mu$ as shown in Definition 1. 
In practice, the wavelength of the dominant spatial mode needs to be longer than the size
of a single cell. This can be tuned by varying the diffusion rate $\mu$
after the finite Turing instability is guaranteed by the
biocircuit structures $A_{\rm circ}$ and $A$.

\par
\smallskip
It is important to guarantee that the dominant spatial mode is 
finite, since it is physically and biologically implausible to achieve 
coordinated spatial patterns of infinite spatial mode at steady state.
The following toy example shows the case where the right-most roots of
$p_{\lambda}(s)=0$ is given in the limit of $\lambda \rightarrow
\infty$. 

\begin{figure}[tb]
\centering
\includegraphics[clip, width=4cm]{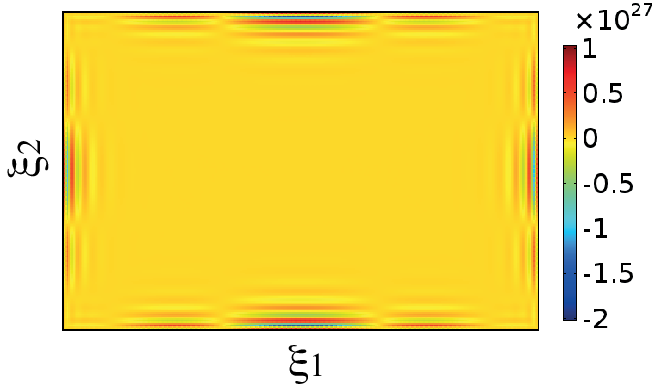}
\includegraphics[clip, width=4.6cm]{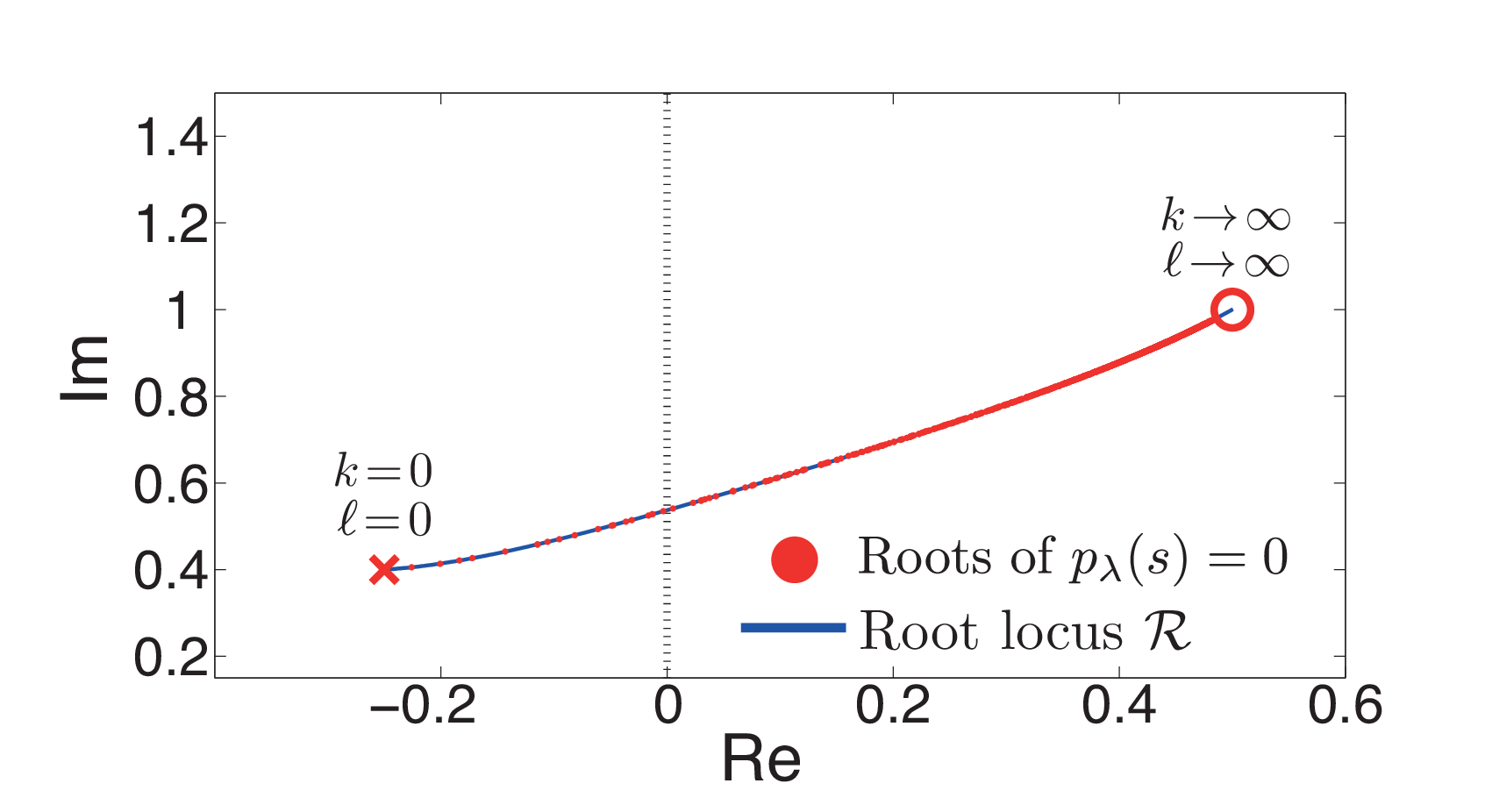}
\caption{(Left) Concentration gradient of the activator $A$. (Right)
 Root locus $\mathcal{R}$ of the cell dynamics $h(s)$.}
\label{type2-eg-fig}
\end{figure}

\par
\smallskip
\noindent
{\bf Example:}
Suppose 
\begin{align}
A = 
\left[
\begin{array}{cc:c}
1.954 & -3.114 & 0 \\
1.0 & -0.9540 & -1.223 \\ \hdashline 
0  & 1.0 & -2.0
\end{array}
\right], 
\end{align}
and $\mu = 0.006$, $L_1 = 3$ and $L_2 = 2$. 
We can calculate the dynamics of a single cell $h(s)$ using
$A$ and the sub-matrix $A_{\rm circ}$ and obtain the root locus
$\mathcal{R}$ of $h(s)$. Figure \ref{type2-eg-fig} (Right) shows the
right-most pole of the root locus $\mathcal{R}$. As $k,\ell \rightarrow
\infty$, or equivalently $\lambda \rightarrow \infty$, the root of
$p_{\lambda}(s)=0$ converges to the root of $n(s) = 0$, or $s_0$. 
This implies that the growth rate of the state $\tilde{{\bm x}}$ is
the largest for the infinite spatial mode, which is not plausible in real
physical and biological systems. In fact, the simulation result in
Fig. \ref{type2-eg-fig} (Left) shows spatial patterns with non-realistic 
high spatial frequency.
The time axis of the simulation, $\tau$, was scaled by $\tau = 200 t$.
The full temporal dynamics is available in a movie format 
\cite{Hori2015Tech}. We used COMSOL$\textregistered $ Multiphysics 4.4 for 
the simulation.

\par
\medskip
It should be noted that the importance of the finite mode instability was not actively discussed in
many classical works \cite{Murray2003, Edelstein-Keshet2005} where the systems are composed
of $n=2$ molecular species and both can diffuse in the spatial domain. 
This is because the dominant spatial mode is always guaranteed to be
finite mode, no matter how the reaction rates are varied.

\subsection{Condition for finite mode Turing instability}
In this section, 
we show that at least three molecular species, 
$\mathcal{M}_1, \mathcal{M}_2$ and $\mathcal{M}_3$, are necessary to 
achieve finite mode Turing instability in molecular communication
networks. This implies that the well-known activator-repressor model \cite{Edelstein-Keshet2005} fails to make spatial
patterns when there is only one diffusible molecule in the system.

\medskip
\noindent
{\bf Assumption 2.~} 
{\it The matrix $A$ in the model (\ref{linear-eq}) is Hurwitz.}

\par
\medskip
This means that the reaction dynamics of a single cell $h(s)$
is stable. As Definition 1 shows, the stability of $h(s)$ is a necessary
condition for finite mode Turing instability, and the existence of
spatial patterns is essentially determined by the condition (ii) of
Definition 1. Hence, by imposing Assumption 2, we hereafter focus on the
condition (ii) of Definition 1. %

\par
\medskip
The following lemma provides the necessary and sufficient condition for
finite mode Turing instability.

\medskip
\noindent
{\bf Lemma 3.~}
{\it 
Consider the biomolecular communication network system
(\ref{nonlinear-eq}) and its linearized system (\ref{linear-eq}). Suppose Assumptions 1 and 2 hold. 
Then, the system (\ref{nonlinear-eq}) is finite mode Turing unstable
around an equilibrium $\bar{{\bm x}}$, if and only if 
either of (a) or (b) holds.

\smallskip
\noindent
{\rm (a)}~ $A_{\rm circ}$ is Hurwitz, and 
$\mathrm{Re}[p_\lambda(j \omega)] = 0,~\mathrm{Im}[p_\lambda( j
\omega)] = 0$ 
for some $\lambda > 0$ and some $\omega \in \mathbb{R}$.

\smallskip
\noindent
{\rm (b)}~ $A_{\rm circ}$ is not Hurwitz, and 
$\mathrm{Re}[p_\lambda(\gamma + j \omega)] = 0,~\mathrm{Im}[p_\lambda(\gamma + j \omega)] = 0$
for some $\lambda > 0$ and some $\omega \in \mathbb{R}$, 
$\gamma := \max_{\nu \in \mathrm{spec}(A_{\rm circ})} \mathrm{Re}[\nu]$
}%

\medskip
\noindent
This lemma is a direct consequence of Definition 1. 
The conditions (a) and (b) are equivalent to (ii) of Definition 1 
with $\gamma = 0$ and $\gamma = \mathrm{Re}[s_0]$, respectively.

\par
\smallskip
Using this lemma, we can derive the following theorem showing that at
least three molecular species are necessary to achieve finite mode Turing
instability. 

\medskip
\noindent
{\bf Theorem 1.~}
{\it 
Consider the biomolecular communication network system
(\ref{nonlinear-eq}) and its linearized system (\ref{linear-eq}). 
Suppose Assumptions 1 and 2 hold. 
Then, the system (\ref{nonlinear-eq}) composed of $n=1$ and $n=2$ molecular species cannot be finite mode Turing unstable. 
}

\medskip
\noindent
The proof of this theorem can be found in Appendix. %
This theorem implies that the activator-repressor-diffuser system illustrated in Section
\ref{ard-sec} is one of the minimal biomolecular communication networks that can exhibit finite mode
Turing instability with $n=3$.

\par
\smallskip
Moreover, the following theorem shows that finite Turing instability
 is always caused by Hopf-Turing bifurcation, implying that not only
coordinated spatial patterns but also temporal oscillations of
concentration gradients are expected at steady state.

\medskip
\noindent
{\bf Theorem 2.~}
{\it 
Consider the biomolecular communication network system
(\ref{nonlinear-eq}) and its linearized system (\ref{linear-eq}).
Suppose the system (\ref{nonlinear-eq}) is finite mode Turing unstable
around an equilibrium $\bar{{\bm x}}$. Then, the right-most pole of the characteristic
polynomial $p_{\lambda}(s) = 0$ is a pair of complex
conjugate.
}%

\medskip
\noindent
{\bf Proof.} 
We prove by contradiction. Suppose the right-most root of $p_{\lambda}(s) = 0$ is a real number
$\sigma$, and it is given when $\lambda = \lambda_0$. That is,
$p_{\lambda_0}(\sigma) = 0$. 
It follows from (i) of Definition 1 that all of the poles of $h(s)$ lie 
in $\mathbb{C}_{-}$. Moreover, (ii) of Definition 1 implies that
$\sigma > {\rm max}(\mathrm{Re}[s_0], 0) \ge 0$, where $s_0$ is a root
of $n(s) = 0$ with the largest real part. Therefore, $\sigma$ is
greater than any real poles and zeros of $h(s)$. 
This, however, contradicts with the property of the root locus that the locus exists on real axis to the
left of an odd number of poles and zeros of $h(s)$~\cite{Levine1996}. 
$\hfill \Box$

\par
\medskip
In the next section, we turn our attention to the synthesis of the minimal molecular communication
networks with $n=3$ molecular species and investigate their properties in detail.

\section{APPLICATION TO THREE-COMPONENT CIRCUIT SYNTHESIS}
\label{threecomp-sec}
In this section, we consider molecular communication networks consisting
of two internal circuit molecules, $\mathcal{M}_1$ and $\mathcal{M}_2$,
and one diffuser $\mathcal{M}_3$. 
For this class of systems, we first derive an analytic condition for
finite mode Turing instability. Using the analytic condition, we
characterize a structural property of biomolecular communication
networks that admit Turing instability.

\par
\smallskip
Let $\alpha_i$ and $\beta_i$ denote the coefficients of the following 
characteristic polynomials. 
\begin{align}
&|sI_3 - A| = s^3 + \alpha_{2} s^{2} +  \alpha_1 s + \alpha_0, \\
&|sI_{2} - A_{\rm circ}| = s^{2} + \beta_{1}s + \beta_0.
\end{align}
Then, a necessary and sufficient condition for finite mode Turing
instability is given by the following theorem.

\medskip
\noindent
{\bf Theorem 3.~}
{\it 
Consider the biomolecular communication network system (\ref{nonlinear-eq})
 composed of $n=3$ molecular species and its linearized system (\ref{linear-eq}). 
Suppose Assumptions 1 and 2. 
The system (\ref{nonlinear-eq}) is finite mode Turing unstable around an equilibrium
$\bar{{\bm x}}$, if and only if the following (A) or (B) is satisfied.
\begin{align}
\mathrm{(A)}&~~
\beta_1 > 0,~\beta_0 > 0, \notag \\
&~~\alpha_1 + \beta_1 \alpha_2 - \beta_0 \le 
-2 \sqrt{\beta_1(\alpha_1 \alpha_2 - \alpha_0)}, \notag \\
\mathrm{(B)}&~~
\beta_1 < 0,~\beta_1^2 - 4 \beta_0 < 0,
 \notag \\ 
&~~-\beta_1^2 + \beta_0 + \beta_1
 \alpha_2 - \alpha_1 > 0.
\notag
\end{align}
}%

\medskip
\noindent
The conditions (A) and (B) in Theorem~3 are equivalent to
(a) and (b) of Lemma~3, respectively. 
The complete proof is shown in Appendix. %
It should be noticed that $\beta_i~(i=0,1)$ are determined from the
parameters of the internal circuit $A_{\rm circ}$, while $\alpha_i~(i=0,1,2)$
depend on the entire system including the diffuser. 
Theorem~3 allows us to further investigate the properties of 
 three-component systems. In what follows, we show that a molecular
 communication network needs to have a certain structural property to admit finite mode Turing instability.

\begin{figure}[tb]
\centering
\includegraphics[clip,width=7cm]{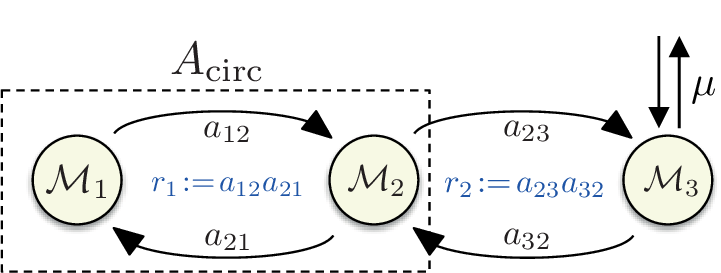}
\caption{Schematic diagram of three-component circuit}
\label{cascade-fig}
\end{figure}

\par
\smallskip
We consider a class of cascaded three-component circuits illustrated in Fig. 
\ref{cascade-fig}. The matrix $A$ is then defined by
\begin{equation}
A = \left[
\begin{array}{cc:c}
a_{11} & a_{12} & 0 \\ 
a_{21} & a_{22} & a_{23} \\ \hdashline
0 & a_{32} & a_{\rm diff}
\end{array}
\right]. 
\end{equation}

\par
\smallskip
The activator-repressor-diffuser network in
Section \ref{ard-sec} is a class of the cascaded circuit with 
$a_{12} < 0$, $a_{21} > 0$, $a_{23} < 0$ and $a_{32} < 0$. 
Note that the signs of $a_{12}, a_{21}, a_{23}$ and $a_{32}$ depend on 
whether the molecule $\mathcal{M}_1, \mathcal{M}_2$ and $\mathcal{M}_3$
are activator or repressor.
The following proposition specifies the combinations of activator and
repressors that admit finite mode Turing instability.

\medskip
\noindent
{\bf Proposition 2.~}
{\it 
Consider the biomolecular communication networks with the cascaded 
structure shown in Fig. \ref{cascade-fig}. Suppose Assumptions 1 and 2 hold.
Let $r_1:=a_{12}a_{21}$ and $r_2:=a_{23}a_{32}$. 
Then, the network admits finite mode Turing instability only if $r_1 < 0$ and $r_2 > 0$. 
}%

\medskip
\noindent
\par
\smallskip
The proof can be found in Appendix. %
This proposition narrows down the design space of biomolecular
communication networks. In other words, the combination of 
the molecules $\mathcal{M}_1$, $\mathcal{M}_2$ and $\mathcal{M}_3$ needs to be either
of (activator, repressor, repressor) or (repressor, activator,
activator) in order for the system to achieve finite mode Turing instability.
It should be noticed that the example shown in Section \ref{ard-sec} is an
example of (activator, repressor, repressor) configuration.

\section{Conclusion}
\label{conc-sec}
In this paper, we have proposed a control theoretic modeling and
analysis framework for biomolecular communication networks
 modeled by reaction-diffusion equations with a single diffusible molecule.
We have first shown that the analysis of the infinite dimensional system
can be simplified by decomposing it into finite dimensional subsystems.
We have then provided a graphical stability analysis tool using the root
locus approach and demonstrated on the novel
activator-repressor-diffuser system, which allows for the formation of 
inhomogeneous concentration gradient over the cell population.
The latter half of the paper has been devoted to analyzing the conditions for finite mode Turing
instability. Using the analytic condition of Turing instability, we have
 shown that the network needs to have at least three molecular species to admit Turing instability. 
Moreover, the in-depth study of three-component circuits provided 
a necessary activation-repression network structure condition for Turing 
instability. 

\par
\smallskip
Toward in vitro and in vivo implementation of such circuits, our future work will be devoted to addressing robustness issues such as
	cell-to-cell variability and parameter uncertainties of	biomolecular
	communication networks. The activator-repressor-diffuser system will also need further study 
to characterize the parameter regions for Turing instability.
\ifCLASSOPTIONcaptionsoff
  \newpage
\fi

\bibliographystyle{ieeetran}
\bibliography{IEEEabrv,IEEEMBMC_Arxiv}

\begin{thebibliography}{10}
\providecommand{\url}[1]{#1}
\csname url@samestyle\endcsname
\providecommand{\newblock}{\relax}
\providecommand{\bibinfo}[2]{#2}
\providecommand{\BIBentrySTDinterwordspacing}{\spaceskip=0pt\relax}
\providecommand{\BIBentryALTinterwordstretchfactor}{4}
\providecommand{\BIBentryALTinterwordspacing}{\spaceskip=\fontdimen2\font plus
\BIBentryALTinterwordstretchfactor\fontdimen3\font minus
  \fontdimen4\font\relax}
\providecommand{\BIBforeignlanguage}[2]{{%
\expandafter\ifx\csname l@#1\endcsname\relax
\typeout{** WARNING: IEEEtran.bst: No hyphenation pattern has been}%
\typeout{** loaded for the language `#1'. Using the pattern for}%
\typeout{** the default language instead.}%
\else
\language=\csname l@#1\endcsname
\fi
#2}}
\providecommand{\BIBdecl}{\relax}
\BIBdecl

\bibitem{Akyildiz2012}
I.~F. Akyildiz, F.~Fekri, R.~Sivakumar, C.~R. Forest, and B.~K. Hammer,
  ``{MONACO}: Fundamentals of molecular nano-communication networks,''
  \emph{IEEE Wireless Communications}, vol.~19, no.~5, pp. 12--18, 2012.

\bibitem{Nakano2014}
T.~Nakano, T.~Suda, Y.~Okaie, M.~J. Moore, and A.~V. Vasilakos, ``Molecular
  communication among biological nanomachines: a layered architecture and
  research issue,'' \emph{IEEE Transactions on NanoBioscience}, vol.~13, no.~3,
  pp. 169--197, 2014.

\bibitem{Danino2010}
T.~Danino, O.~Mondrag\'{o}n-Palomino, L.~Tsimring, and J.~Hasty, ``A
  synchronized quorum of genetic clocks,'' \emph{Nature}, vol. 463, no. 7279,
  pp. 326--330, 2010.

\bibitem{You2004}
L.~You, R.~S.~C. III, R.~Weiss, and F.~H. Arnold, ``Programmed population
  control by cell-cell communication and regulated killing,'' \emph{Nature},
  vol. 428, no. 6985, pp. 868--871, 2004.

\bibitem{Balagadde2008}
F.~K. Balagadd\'{e}, H.~Song, J.~Ozaki, C.~H. Collins, M.~Barnet, F.~H. Arnold,
  S.~R. Quake, and L.~You, ``A synthetic {Escherichia} coli
  predator^^e2^^80^^93prey ecosystem,'' \emph{Molecular Systems Biology},
  vol.~4, no.~1, 2008.

\bibitem{Tabor2009}
J.~J. Tabor, H.~Salis, Z.~B. Simpson, A.~A. Chevalier, A.~Levskaya, E.~M.
  Marcotte, C.~A. Voigt, and A.~D. Ellington, ``A synthetic genetic edge
  detection program,'' \emph{Cell}, vol. 137, no.~7, pp. 1272--1281, 2009.

\bibitem{Tamsir2011}
A.~Tamsir, J.~J. Tabor, and C.~A. Voigt, ``Robust multicellular computing using
  genetically encoded nor gates and chemical ‘wires’,'' \emph{Nature}, vol.
  469, no. 7329, pp. 212--215, 2011.

\bibitem{Regot2011}
S.~Regot, J.~Macia, N.~Conde, K.~Furukawa, J.~Kjell\'{e}n, T.~Peeters,
  S.~Hohmann, E.~de~Nadal, F.~Posas, and R.~Sol\'{e}, ``Distributed biological
  computation with multicellular engineered networks,'' \emph{Nature}, vol.
  469, no. 7329, pp. 207--211, 2011.

\bibitem{Basu2005}
S.~Basu, Y.~Gerchman, C.~H. Collins, F.~H. Arnold, and R.~Weiss, ``A synthetic
  multicellular system for programmed pattern formation,'' \emph{Nature}, vol.
  434, no. 7037, pp. 1130--1134, 2005.

\bibitem{Liu2011}
C.~Liu, X.~Fu, L.~Liu, X.~Ren, C.~K. Chau, S.~Li, L.~Xiang, H.~Zeng, G.~Chen,
  L.-H. Tang, P.~Lenz, X.~Cui, W.~Huang, T.~Hwa, and J.-D. Huang, ``Sequential
  establishment of stripe patterns in an expanding cell population,''
  \emph{Science}, vol. 334, no. 6053, pp. 238--241, 2011.

\bibitem{Koch1994}
A.~Koch and H.~Meinhardt, ``Biological pattern formation: from basic mechanisms
  to complex structures,'' \emph{Reviews of Modern Physics}, vol.~66, no.~4,
  pp. 1481--1507, 1994.

\bibitem{Kondo2010}
S.~Kondo and T.~Miura, ``Reaction-diffusion model as a framework for
  understanding biological pattern formation,'' \emph{Science}, vol. 329, no.
  5999, pp. 1616--1620, 2010.

\bibitem{Hsia2012}
J.~Hsia, W.~J. Holtz, D.~C. Huang, M.~Arcak, and M.~M. Maharbiz, ``A feedback
  quenched oscillator produces {Turing} patterning with one diffuser,''
  \emph{PLOS Computational Biology}, vol.~8, no.~1, 2012, e1002331.

\bibitem{Edelstein-Keshet2005}
L.~Edelstein-Keshet, \emph{Mathematical models in biology}.\hskip 1em plus
  0.5em minus 0.4em\relax SIAM, 2005.

\bibitem{Turing1952}
A.~M. Turing, ``The chemical basis of morphogenesis,'' \emph{Philosophicals
  Transactions of the Royal Society of London B}, vol. 237, no. 641, pp.
  37--72, 1952.

\bibitem{Gierer1972}
A.~Gierer and H.~Meinhardt, ``A theory of biological pattern formation,''
  \emph{Kybernetik}, vol.~12, no.~1, pp. 30--39, 1972.

\bibitem{Othmer2009}
H.~G. Othmer, K.~Painter, D.~Umulis, and C.~Xue, ``The intersection of theory
  and application in elucidating pattern formation in developmental biology,''
  \emph{Mathematical Modelling of Natural Phenomena}, vol.~4, no.~4, pp. 3--82,
  2009.

\bibitem{Levine1996}
W.~S. Levine, \emph{The Control Handbook}.\hskip 1em plus 0.5em minus
  0.4em\relax CRC Press, 1996.

\bibitem{MiyazakoCDC}
H.~Miyazako, Y.~Hori, and S.~Hara, ``Turing instability in reaction-diffusion
  systems with a single diffuser: characterization based on root locus,'' in
  \emph{Proceedings of IEEE Conference on Decision and Control}, 2013, pp.
  2671--2676.

\bibitem{HoriSICE2014}
Y.~Hori, S.~Kumagai, and S.~Hara, ``Network structure for {Turing}
  instabilizability in reaction-diffusion systems with one diffuser: a case
  study for three-gene networks,'' in \emph{Proceedings of SICE Annual
  Conference}, 2014, pp. 892--895.

\bibitem{Miyazako2013}
H.~Miyazako, Y.~Hori, and S.~Hara, ``The analysis of {Turing} instability in
  reaction-diffusion systems using a single diffuser,'' \emph{Transactions of
  the Society of Instrument and Control Engineers}, vol.~49, no.~12, pp.
  1164--1171, 2013, (in Japanese).

\bibitem{Arcak2011}
M.~Arcak, ``Certifying spatially uniform behavior in reaction-diffusion pde and
  compartmental {ODE} systems,'' \emph{Automatica}, vol.~47, no.~6, pp.
  1219--1229, 2011.

\bibitem{Gardner2000}
T.~S. Gardner, C.~R. Cantor, and J.~J. Collins, ``Construction of a genetic
  toggle switch in escherichia coli,'' \emph{Nature}, vol. 403, no. 6767, pp.
  339--342, 2000.

\bibitem{Elowitz2000}
M.~B. Elowitz and S.~Leibler, ``A synthetic oscillatory network of
  transcriptional regulators,'' \emph{Nature}, vol. 403, no. 6767, pp.
  335--338, 2000.

\bibitem{Moon2012}
T.-S. Moon, C.~Lou, A.~Tamsir, B.~C. Stanton, and C.~A. Voigt, ``Genetic
  programs constructed from layered logic gates in single cells,''
  \emph{Nature}, vol. 491, no. 7423, pp. 249--253, 2012.

\bibitem{Morgan1989}
J.~Morgan, ``Global existence for semilinear parabolic systems,'' \emph{SIAM
  Journal on Mathematical Analysis}, vol.~20, no.~5, pp. 1128--1144, 1989.

\bibitem{Smoller1994}
J.~Smoller, \emph{Shock waves and reaction-diffusion equations}.\hskip 1em plus
  0.5em minus 0.4em\relax Springer-Verlag, 1994.

\bibitem{Hara2009}
S.~Hara, T.~Hayakawa, and H.~Sugata, ``{LTI} systems with generalized frequency
  variables: A unified framework for homogeneous multi-agent dynamical
  systems,'' \emph{SICE Journal of Control, Measurement and System
  Integration}, vol.~2, no.~5, pp. 299--306, 2009.

\bibitem{Hara2014}
S.~Hara, H.~Tanaka, and T.~Iwasaki, ``Stability analysis of systems with
  generalized frequency variables,'' \emph{IEEE Transactions on Automatic
  Control}, vol.~59, no.~2, pp. 313--326, 2014.

\bibitem{Nakamura2014}
T.~Nakamura, Y.~Hori, and S.~Hara, ``Hierarchical modeling and local stability
  analysis for repressilators coupled by quorum sensing,'' \emph{SICE Journal
  of Control, Measurement, and System Integration}, vol.~7, no.~3, pp.
  133--140, 2014.

\bibitem{Hespanha2009}
J.~P. Hespanha, \emph{Linear systems theory}.\hskip 1em plus 0.5em minus
  0.4em\relax Princeton University Press, 2009.

\bibitem{Strauss2008}
W.~A. Strauss, \emph{Partial differential equations: An introduction},
  2nd~ed.\hskip 1em plus 0.5em minus 0.4em\relax John Wiley, 2008.

\bibitem{Holland1978}
R.~G. Casten and C.~J. Holland, ``Instability results for reaction diffusion
  equations with neumann boundary conditions,'' \emph{Journal of Differential
  Equations}, vol.~27, no.~2, pp. 266--273, 1978.

\bibitem{Hori2015Tech}
Y.~Hori, H.~Miyazako, S.~Kumagai, and S.~Hara, ``Coordinated spatial pattern
  formation in biomolecular communication networks,'' 2015, arXiv:1504.06045v3
  (available at http://arxiv.org/abs/1504.06045).

\bibitem{Rhodius2013}
V.~A. Rhodius, T.~H. Segall-Shapiro, B.~D. Sharon, A.~Ghodasara, E.~Orlova,
  H.~Tabakh, D.~H. Burkhardt, K.~Clancy, T.~C. Peterson, C.~A. Gross, and C.~A.
  Voigt, ``Design of orthogonal genetic switches based on a crosstalk map of
  $\sigma$s, anti-$\sigma$s, and promoters,'' \emph{Molecular Systems Biology},
  vol.~9, p. 702, 2013.

\bibitem{Stanton2014}
B.~C. Stanton, A.~A.~K. Nielsen, A.~Tamsir, K.~Clancy, T.~Peterson, and C.~A.
  Voigt, ``Genomic mining of prokaryotic repressors for orthogonal logic
  gates,'' \emph{Nature Chemical Biology}, vol.~10, pp. 99--105, 2014.

\bibitem{Mutalik2013}
V.~K. Mutalik, J.~C. Guimaraes, G.~Cambray, C.~Lam, M.~J. Christoffersen, Q.-A.
  Mai, A.~B. Tran, M.~Paull, J.~D. Keasling, A.~P. Arkin, and D.~Endy,
  ``Precise and reliable gene expression via standard transcription and
  translation initiation elements,'' \emph{Nature Methods}, vol.~10, pp.
  354--360, 2013.

\bibitem{Henrike2015}
H.~Niederholtmeyer, Z.~Sun, Y.~Hori, E.~Yeung, A.~Verpoorte, R.~M. Murray, and
  S.~J. Maerkl, ``Rapid cell-free forward engineering of novel genetic ring
  oscillators,'' \emph{eLife}, 2015, (accepted).

\bibitem{Hsia2012CDC}
J.~Hsia, W.~J. Holtz, M.~M. Maharbiz, and M.~Arcak, ``New architecture for
  patterning gene expression using zinc finger proteins and small {RNA}s,'' in
  \emph{Proceedings of IEEE Conference on Decision and Control}, 2012, pp.
  1633--1638.

\bibitem{Murray2003}
J.~Murray, \emph{Mathematical Biology II}, 3rd~ed.\hskip 1em plus 0.5em minus
  0.4em\relax Springer, 2003.

\end{thebibliography}

\begin{appendix}
\subsection{Proof of Lemma 1} 
\label{prop2-proof}
Using the definition (\ref{h-def}), we can write $h(s)$ as 
\begin{align}
h(s) = \frac{{\bm e_n}^T \mathrm{adj}(sI_n - A) {\bm e_n}}{|sI_n - A|}, 
\end{align}
where $\mathrm{adj}(X)$ denotes the adjugate matrix of $X$.
It follows from the definition of the adjugate matrix that 
the $(n,n)$-th entry of $\mathrm{adj}(sI_n - A)$ is $|sI_{n-1} - A_{\rm
circ}|$, and thus the numerator of $h(s)$ can be written as ${\bm e}_n^T \mathrm{adj}(sI_n - A) {\bm e}_n = |sI_{n-1} - A_{\rm
circ}|$.~$\hfill \Box$

\subsection{Proof of Theorem 1}
\label{theo1-proof}
For $n=1$, it follows that the molecular communication network is 
finite mode Turing unstable if and only if the characteristic polynomial $p_{\lambda}(s) = s - a_{\rm diff} +
\lambda$ satisfies (a) of Lemma 3. We can easily see that 
$\mathrm{Re}[p_\lambda(j\omega)] = 0$ cannot be satisfied for all
$\lambda > 0$ and $\omega \in \mathbb{R}$, since $a_{\rm diff} < 0$.

\par
\smallskip
For $n=2$, we define coefficient $\alpha_0, \alpha_1$ and $\beta_0$ by $|sI_2 - A| = s^2 + \alpha_1 s + \alpha_0$ and 
$|sI_2 - A_{\rm circ}| = s + \beta_0$. 
In what follows, %
we show that both (a) and (b) of Lemma 3 cannot be satisfied.

\smallskip
\noindent
{\bf Case 1: $A_{\rm circ}$ is Hurwitz:}~
It follows that $\mathrm{Im}[p_{\lambda}(j\omega)] = \omega
(\alpha_1 + \lambda)$ for all $\lambda > 0$ and $\omega \neq 0$, since $\alpha_1 > 0$ holds when $A$ is Hurwtiz. 
For $\omega = 0$, $\mathrm{Re}[p_{\lambda}(0)] = \alpha_0 + \lambda
\beta_0 \neq 0$ holds for all $\lambda > 0$ because of $\alpha_0 > 0$
and $\beta_0 > 0$. Thus, the condition (a) in Lemma 3 cannot be satisfied. 

\smallskip
\noindent
{\bf Case 2: $A_{\rm circ}$ is not Hurwitz:}~
It follows that $\mathrm{Im}[p_{\lambda}(\gamma + j\omega)] =
\omega(2 \gamma + \alpha_1 + \lambda)$, where $\gamma = \beta_0 (\ge 0)$ from
the definition.
Thus, $\mathrm{Im}[(\lambda, \gamma + j\omega)] \neq 0$ for all
$\lambda > 0$ and $\omega \neq 0$, since $\alpha_1 > 0$ and $\gamma \ge 0$.
When $\omega = 0$, $\mathrm{Re}[p_{\lambda}(\gamma)] = \gamma^2 + 
\gamma(\alpha_1 + \lambda) + \alpha_0 + \lambda \beta_0 > 0$ hold, and
$\mathrm{Re}[p_{\lambda}(\gamma)] \neq 0$ 
 hold for all $\lambda > 0$. This completes the proof.~$\hfill \Box$

\subsection{Proof of Theorem 3}
\label{theo3-proof}
We prove the theorem by calculating conditions (a) and 
(b) of Lemma 3 for $n=3$.

\par
\smallskip
In what follows, we derive the conditions (A) and (B) by
considering the case where $A_{\rm circ}$ is Hurwitz and
not Hurwitz, respectively.
We can easily verify that neither (a) nor (b) of Lemma 3
can be satisfied when $\omega = 0$. 
Hence, we hereafter show the proof for the case of $\omega \neq 0$.

\smallskip
\noindent
{\bf Case 1: $A_{\rm circ}$ is Hurwitz:~}
A necessary condition for $A_{\rm circ}$ being
Hurwitz is 
\begin{align}
\beta_1 > 0~\mathrm{and}~\beta_0 > 0.
\label{case1-tildeA-eq}
\end{align}
Moreover, $p_{\lambda}(j\omega) = 0$ implies that 
\begin{align}
&\mathrm{Re}[p_\lambda(j\omega)] = -\alpha_2 \omega^2 + \alpha_0 +
 \lambda(-\omega^2 + \beta_0) = 0 \label{case1-real-eq}\\
&\mathrm{Im}[p_\lambda(j\omega)] = -\omega\{\omega^2 - (\alpha_1 +
 \lambda \beta_1)\} = 0.\label{case1-imag-eq}
\end{align}
It follows from (\ref{case1-imag-eq}) that $\omega^2 = \alpha_1 +
\lambda \beta_1 (> 0)$, since $\omega \neq 0$. 
We can then eliminate $\omega^2$ from (\ref{case1-real-eq}), and we have
the following quadratic equation of $\lambda$.
\begin{align}
\beta_1 \lambda^2 + (\alpha_1 + \beta_1 \alpha_2 -
 \beta_0)
\lambda + \alpha_1 \alpha_2 - \alpha_0 = 0.
\label{case1-lambda-eq}
\end{align}
Since $\beta_1 > 0$ holds from (\ref{case1-tildeA-eq}), and $\alpha_1
\alpha_2 - \alpha_0 > 0$ holds from Assumption 2, a necessary
and sufficient condition for (\ref{case1-lambda-eq}) having a positive
real solution is given by the following (C1) and (C2).

\smallskip
\noindent
(C1) The determinant of (\ref{case1-lambda-eq}) is non-negative. That
is, 
\begin{align}
((\alpha_1 + \beta_1) \alpha_2 - \beta_0^2)^2 -
 4\beta_1(\alpha_1 \alpha_2 - \alpha_0) \ge 0.
\label{C1-eq}
\end{align}

\smallskip
\noindent
(C2) The coefficient of $\lambda$ in (\ref{case1-lambda-eq}) is
negative. That is, 
\begin{align}
\alpha_1 + \beta_1 \alpha_2 - \beta_0 < 0.
\label{C2-eq}
\end{align}

\par
\smallskip
Summarizing (\ref{C1-eq}) and (\ref{C2-eq}), we have 
\begin{align}
\alpha_1 + \beta_1 \alpha_2 - \beta_0 \le 
-2 \sqrt{\beta_1 (\alpha_1 \alpha_2 - \alpha_0)}.
\label{case1-last-eq}
\end{align}
The condition (A) is obtained from (\ref{case1-tildeA-eq}) and 
(\ref{case1-last-eq}).

\medskip
\noindent
{\bf Case 2: $A_{\rm circ}$ is not Hurwitz~}
A necessary condition for $A_{\rm circ}$ not being Hurwitz is 
\begin{align}
\beta_1 \le 0~\mathrm{or}~\beta_0 \le 0.
\label{case2-tildeA-eq}
\end{align}
Let $\gamma \pm j \eta~(\gamma, \eta \ge 0)$ denote a pair of 
eigenvalues of $A_{\rm circ}$ with the largest real part. 
Then, $|(\gamma + j\eta)I_2 - A_{\rm circ}| = 0$ implies 
\begin{align}
&\gamma^2 + \beta_1 \gamma + \beta_0 - \eta^2 = 0
\label{case2-tildeA-det-eq1}
\\
&(2 \gamma + \beta_1) \eta = 0.
\label{case2-tildeA-det-eq2}
\end{align}

\par
\smallskip
In what follows, we first show that $p_\lambda(\gamma + j\eta) \neq 0$ for all $\lambda > 0$ 
and $\omega$ when $\eta = 0$.
It follows from $\mathrm{Im}[p_\lambda(\gamma + j\omega)] = 0$ that 
\begin{align}
\omega^2 = 3 \gamma^2 + 2 \alpha_2 \gamma + \alpha_1 + \lambda(2\gamma +
 \beta_1).
\end{align}
Substituting this into $\mathrm{Re}[p_\lambda(\gamma + j\omega)] = 0$
yields the following equation of $\lambda$.
\begin{align}
(&2\gamma + \beta_1) \lambda^2 + \{
9\gamma^2 + (4 \alpha_2 + 3 \beta_1)\gamma + \alpha_1 +
 \beta_1 \alpha_2
\}\lambda \notag \\
&+
\{
8\gamma^3 + 8\alpha_2 \gamma^2 + 2(\alpha_1 + \alpha_2^2)\gamma
 + (\alpha_1 \alpha_2 - \alpha_0)
\} = 0.
\label{case2-gamma0-eq}
\end{align}
It should be noted that the coefficient of $\lambda^2$ is non-negative, or 
$2 \gamma + \beta_1 \ge 0$, since $\gamma$ is the 
largest real root of $s^2 + \beta_1 s + \beta_0 = 0$
from the definition.
Moreover, the coefficients of $\lambda$ and the constant terms of (\ref{case2-gamma0-eq}) 
are also positive because $\alpha_0 > 0$, $\alpha_2 > 0$ and $\alpha_1
\alpha_2 - \alpha_0 > 0$ from Assumption 2, and (\ref{case2-tildeA-eq}) holds.
Thus, $p_\lambda(\gamma + j\eta) \neq 0$ for all $\lambda > 0$ 
and $\omega$ when $\eta = 0$.

\par
\smallskip
When $\eta \neq 0$, the determinant of $|sI_2 - A_{\rm circ}| = 0$ is
negative, thus we have 
\begin{align}
\beta_1^2 - 4 \beta_0 < 0.
\label{case2-det-eq}
\end{align}
Moreover, $\mathrm{Im}[p_\lambda(\gamma + j\omega)] = 0$ implies 
\begin{align}
\omega^2 = 3 \gamma^2 + 2\alpha_2 \beta + \alpha_1.
\end{align}
Substituting this into $\mathrm{Re}[p_\lambda(\gamma + j\omega)] = 0$ yields
\begin{align}
\{\gamma^2 -& (3\beta^2 + 2 \alpha_2 \beta + \alpha_1)\}\lambda 
 - 
\{
8\gamma^3 + 8\alpha_2 \gamma^2 + \notag \\
&2(\alpha_1 +
 \alpha_2^2)\gamma 
 + 
(\alpha_1 \alpha_2 - \alpha_0)
\}
 = 0.
\label{case2-gammanon0-eq}
\end{align}
Note that the constant term of (\ref{case2-gammanon0-eq}) is negative 
 because of Assumption 2 and $\gamma > 0$.
Thus, (\ref{case2-gammanon0-eq}) has a positive real root if and only if 
\begin{align}
\eta^2 - (3\gamma^2 + 2 \alpha_2 \gamma + \alpha_1) > 0.
\label{case2-last-ineq}
\end{align}
The condition (B) is obtained from (\ref{case2-tildeA-eq}), 
(\ref{case2-tildeA-det-eq1}), (\ref{case2-tildeA-det-eq2}), 
(\ref{case2-det-eq}) and (\ref{case2-last-ineq}).
$\hfill \Box$

\subsection{Proof of Proposition 2}
\label{prop3-proof}
We first show $r_1 < 0$. 
Suppose (A) of Theorem 3 holds. 
It follows from $\beta_1 = -(a_{11} + a_{22}) > 0$ of (A) that either or both
of $a_{11}$ and $a_{22}$ need to be negative. 
It follows that $r_1 = a_{12}a_{21} < 0$ when $a_{11}$ and $a_{22}$
have opposite signs, hence we consider the case of $a_{11} < 0$ and
$a_{22} < 0$ in the following.

\par
\smallskip
We first note that $\alpha_1\alpha_2 - \alpha_0 > 0$ holds from
Assumption 2, and it can be equivalently written as
\begin{align}
(a_{11} &+ a_{22})r_1 + (a_{22} + a_{\rm diff})r_2 \notag \\
& - (a_{11} + a_{22})(a_{22} + a_{\rm diff})(a_{\rm diff} + a_{11}) > 0. \label{eq: 1-2}
\end{align}
Dividing (\ref{eq: 1-2}) by $a_{11} + a_{22} (< 0)$, we have 
\begin{align}
r_1 <(a_{22} + a_{\rm diff})(a_{\rm diff} + a_{11}) - \frac{a_{22} +
 a_{\rm diff}}{a_{11} + a_{22}}r_2. \label{eq: caseA-1-1}
\end{align}
On the other hand, $r_2$ satisfies the following inequality from 
$\alpha_1 + \beta_1 \alpha_2 - \beta_0 \le -2 \sqrt{\beta_1 (\alpha_1
\alpha_2 - \alpha_0)} \le 0$ of (A). %
\begin{eqnarray}
(a_{11} + a_{22})(a_{11} + a_{22} + 2a_{\rm diff}) - r_2 \le 0. \label{eq: caseA-1-2}
\end{eqnarray}
Since $a_{11} < 0, a_{22} < 0$ and $a_{\rm diff} < 0$, the second term on the right
side of \eqref{eq: caseA-1-1} satisfies 
\begin{eqnarray}
-\frac{a_{22} + a_{\rm diff}}{a_{11} + a_{22}}r_2 < -(a_{22} + a_{\rm
 diff})(a_{11} +a_{22} + 2a_{\rm diff}). \label{eq: caseA-1-3}
\end{eqnarray}
Summerizing \eqref{eq: caseA-1-1} and \eqref{eq: caseA-1-3}, we have
\begin{align}
r_1 &< (a_{22} + a_{\rm diff})(a_{\rm diff} + a_{11}) - \frac{a_{22} +
 a_{\rm diff}}{a_{11} + a_{22}}r_2 \notag \\
    &< (a_{22} + a_{\rm diff})(a_{\rm diff} + a_{11}) \notag \\
&~~~~~ - (a_{22} + a_{\rm
		 diff})(a_{11} +a_{22} + 2a_{\rm diff}) \notag \\
    &= -(a_{22} + a_{\rm diff})^2 < 0. 
\end{align}

\par
\smallskip
Suppose (II-B) of Theorem 3 holds. It then follows from 
$\beta_1^2 - 4\beta_0 < 0$ of (B) that 
\begin{align}
(a_{11}+ a_{22})^2 - 4(a_{11} a_{22} - r_1) = (a_{11} - a_{22})^2 + 4
 r_1 < 0. 
\notag
\end{align}
This concludes $r_1 < 0$.

\par
\smallskip
Next, we show that $r_2 > 0$ is necessary. Suppose (B) of Theorem 3
holds. Then, the direct calculation leads to $-\beta_1^2 + \beta_0 + \beta_1
\alpha_2 - \alpha_1 = r_2 > 0$. In what follows, we consider the case where (A)
of Theorem 3 holds.

\par
\smallskip
It follows from $a_{\rm diff} < 0$ and $\beta_1= -(a_1 + a_2) > 0$ of
(A) that $(a_1 + a_2 + 2a_3) < 0$. 
Then, $\alpha_1 + \beta_1 \alpha_2 - \beta_0 \le -2 \sqrt{\beta_1 (\alpha_1
\alpha_2 - \alpha_0)} \le 0$ of (A) implies (\ref{eq: caseA-1-2}).
Thus, $r_2 > (a_1 + a_2)(a_1 + a_2 + 2a_3) > 0$.~$\hfill \Box$

\end{appendix}
\end{document}